\documentclass[conference]{IEEEtran}
\usepackage{algorithm}
\usepackage{algorithmic}
\IEEEoverridecommandlockouts
\usepackage{cite}
\usepackage{amsmath,amssymb,amsfonts}
\usepackage{algorithmic}
\usepackage{graphicx}
\usepackage{textcomp}
\usepackage{xcolor}
\usepackage{booktabs}

\def\BibTeX{{\rm B\kern-.05em{\sc i\kern-.025em b}\kern-.08em
    T\kern-.1667em\lower.7ex\hbox{E}\kern-.125emX}}

\begin{document}

\title{HunterAgent: Neuro-Symbolic Attack Trace Reconstruction under Anti-Forensics}

\author{
  Guangze Zhao$^{1,2,4}$,
  Yongzheng Zhang$^{1}$,
  Weilin Gai$^{3,4}$,
  Hongri Liu$^{2}$,
  Yuliang Wei$^{2}$,
  Bailing Wang$^{2,5,6,\dagger}$
  \thanks{
  $^1$Faculty of Computing, Harbin Institute of Technology, China.
  $^2$School of Computer Science and Technology, Harbin
Institute of Technology (Weihai), China.
  $^3$National Computer Network Emergency Response Technical Team/Coordination Center of China (CNCERT/CC).
  $^4$Chang An Communication Technology Co., Ltd., China.
  $^5$Harbin Institute of Technology (Weihai) Qingdao Research Institute, China.
  $^6$Shandong Key Laboratory of Industrial Network Security, China.
  $^{\dagger}$Corresponding author: Bailing Wang (\texttt{wbl@hit.edu.cn}).
  This work was supported by National Natural Science Foundation of China (NSFC) (Grant No.~62272129), Key R\&D Program of Shandong Province (Grant No.~2023CXPT065).
  }
}

% --- 修改部分结束 ---

\maketitle

\begin{abstract}
Modern alert-triage systems have significantly reduced the burden on
Security Operations Centers (SOCs) by filtering false positives, yet
flagging a high-risk alert is only the beginning of incident response. The
subsequent phase---\textit{threat hunting and investigation}---requires
analysts to reconstruct the causal attack chain across heterogeneous,
partially corrupted logs. Against Advanced Persistent Threats (APTs)
employing targeted anti-forensics (parent-PID spoofing, log wiping,
fileless execution), deterministic provenance graphs sever into
disjoint subgraphs and stall. Deploying autonomous Large Language Model
(LLM) agents to bridge the gap introduces a different failure: treating
retrospective tracing as open-ended text generation, unconstrained
agents fabricate causal links that violate OS physics, producing
narratives that are fluent but inadmissible for forensic audit.
We propose \textbf{HunterAgent}, a neuro-symbolic threat-hunting
framework that reframes trace reconstruction as a cost-bounded
heuristic graph search under partial observability. HunterAgent employs
an \textit{Asymmetric Generator--Verifier} pipeline: the LLM acts
strictly as a semantic hypothesis proposer within a typed ontology,
while a deterministic Verifier grounds each hypothesis through
identifier-level collisions on \emph{surviving orthogonal telemetry}
(network 5-tuples, file inodes, PPID chains, ETW callbacks). To resolve
completely severed traces, we score each unverified hop by a calibrated
additive deviation cost combining contextual semantic divergence and an
empirical OS log-normal temporal potential, with schema violations
hard-pruned rather than blended into the cost. A length-discounted
cumulative epistemic budget---calibrated to the 99-th percentile of
benign leave-one-edge-out reconstruction cost---bounds inferential
drift and forces graceful halting in lieu of fabrication.
Under strict Leave-One-Family-Out (LOFO) cross-validation on three
public benchmarks (DARPA TC E3 / OpTC / ATLAS) plus an in-house
40-trace multi-source benchmark (APT-Eval-Trace), HunterAgent
attains a mean \textbf{86.1\% F1} across the four datasets
(\textbf{86.8\% on the in-house D4}; \textbf{85.8\% averaged on the
three public datasets}), outperforming the strongest agentic
baseline by \textbf{+26.7 F1 mean (+28.7 on D4)} and the strongest
non-agentic baseline (KAIROS) by \textbf{+17.1 F1 mean (+20.3 on
D4)}, while suppressing the path-level hallucination rate from
61.5\% to \textbf{6.4\%}.
Under 70\% targeted log wiping, recall degrades but precision
remains $\ge$84\%, with 95.7\% of investigations halting safely
rather than fabricating chains---preserving audit-grade forensic
integrity. All results hold under the realistic
\emph{bounded-evidence assumption}: at least one orthogonal telemetry source survives at the relevant OS layer.
\end{abstract}


\section{Introduction}

Modern Security Operations Centers (SOCs) operate under constant siege
from the sheer volume and heterogeneity of enterprise telemetry
\cite{hassan2019nodoze, milajerdi2019holmes}. Recent advances in alert
triage systems~\cite{pei2025selfprompt} have successfully mitigated initial alert fatigue,
efficiently isolating genuine high-priority anchors from a sea of false
positives. However, triage only answers \emph{whether} a system is under
attack. The true operational bottleneck lies in the subsequent phase---
\textit{threat hunting and investigation}---which requires analysts to
reconstruct the \emph{how}, \emph{when}, and \emph{where} of the entire
attack lifecycle (the provenance) across disconnected, partially
corrupted data silos~\cite{bhattarai2023prov2vec}.

\begin{figure}[t]
    \centering
    \includegraphics[width=\columnwidth]{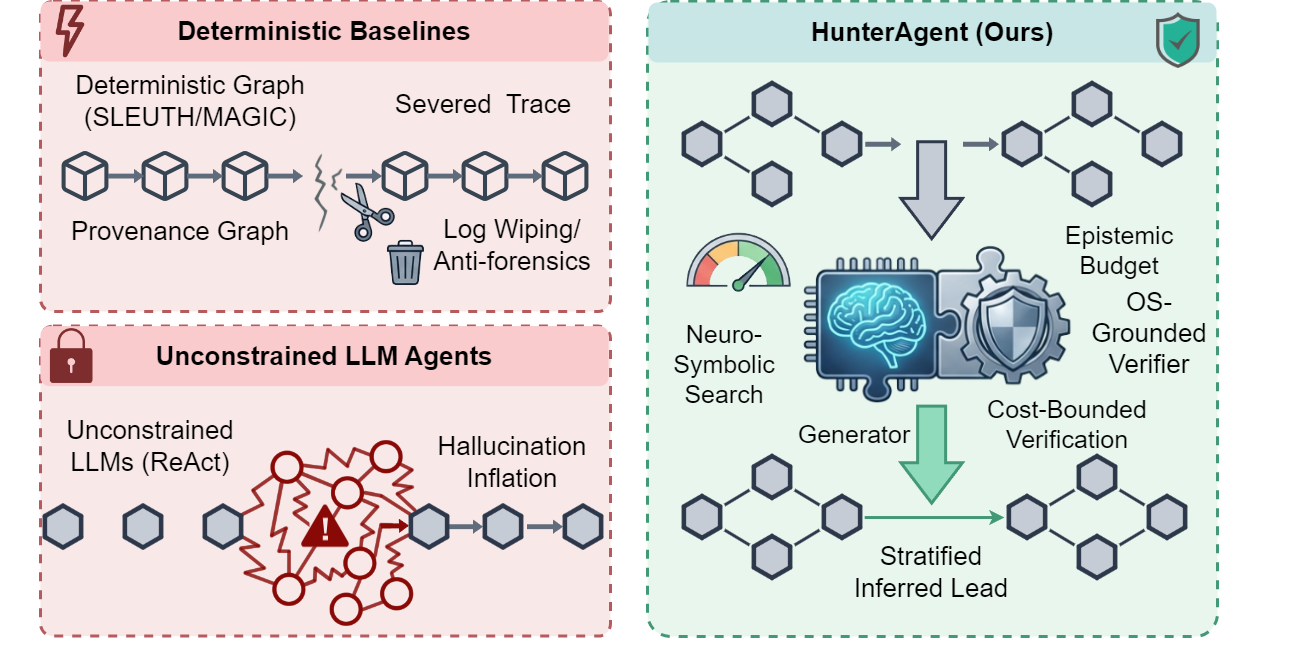}
    \caption{Conceptual comparison of threat-hunting paradigms.
    Deterministic graphs sever under anti-forensics; unconstrained LLMs
    inflate hallucinations; HunterAgent uses a constrained
    neuro-symbolic search bounded by an empirical epistemic budget to
    safely bridge gaps---and to halt explicitly when evidence is
    insufficient.}
    \label{fig:intro_comparison}
\end{figure}

Automated threat attribution today relies predominantly on deterministic
provenance graph analysis~\cite{wang2020you} (e.g., SLEUTH
\cite{hossain2017sleuth}, MAGIC \cite{hassan2020tactical}). Such
rule-based systems are highly reliable against uninterrupted malware
execution, but become brittle under \textit{targeted} anti-forensics.
Advanced Persistent Threats (APTs) routinely clear specific event logs,
spoof parent process IDs, or execute filelessly in
memory~\cite{goyal2023sometimes, fan2026graph}. The removal of even a
single causal record fractures the deterministic graph into disjoint
subgraphs, halting automated traversal and offloading the burden of
bridging the chasm onto human analysts armed only with historical
threat intelligence.

Autonomous LLM agents (e.g., ReAct-based
architectures~\cite{yao2022react}) bring profound zero-shot semantic
reasoning to security operations~\cite{hasanov2024application,
ali2025beyond, ma2025heuristic, jia2025omnisafebench,cheng2024reinforcement,cao2025agr,duan2025oyster,cheng2025inverse}, and at first glance appear well-suited to bridge these
anti-forensic gaps. Yet deploying them directly for retrospective trace
reconstruction introduces a different, equally severe failure mode.
Unconstrained agents treat the task as open-ended text generation,
without formal verification boundaries or OS-level temporal grounding,
and consequently fabricate causal links that satisfy narrative
coherence but violate operating-system physics (e.g., an unprivileged
web worker "directly injecting" into a PPL-protected service, or a
payload executing before its parent finishes writing it to disk).
Because forensic outputs must be defensible in a SOC audit, even a
single such hallucinated edge poisons the entire investigation report~\cite{pang2026steering}.
Existing attempts at constrained decoding or graph-based
RAG~\cite{edge2024local, hakim2025neuro, eckhoff2025experimenting,cheng2025ecoalign}
mitigate but do not eliminate this risk: they constrain \emph{syntax},
not \emph{physics}.

As illustrated in Fig.~\ref{fig:intro_comparison}, we present
\textbf{HunterAgent}, which reframes automated post-compromise hunting
from fragile deterministic graph matching---\emph{and} from
unconstrained LLM generation---into a score-based heuristic search
under partial observability. HunterAgent is built on an
\textit{Asymmetric Generator--Verifier} architecture: the LLM acts
strictly as a semantic hypothesis proposer, restricted by a typed
system ontology and constrained JSON grammar to emit only the
\emph{semantic payload} of a candidate node; a deterministic, OS-grounded
Verifier then attempts to ground each hypothesis through identifier-level
collisions against \emph{surviving orthogonal telemetry} (e.g., shared
network 5-tuples, file inodes, PPID chains, ETW callbacks). This
decoupling lets us preserve LLM generalisation while delegating
correctness to a deterministic gatekeeper that an analyst---or a
court---can audit.

When no surviving telemetry can ground a hypothesis, the trace remains
severed and the search must reason counterfactually. HunterAgent
expands the causal graph through a \textit{Cost-Bounded Beam Search}
that frames missing-log inference as a cost-minimisation problem. The
deviation cost $C_{dev}$ is a calibrated, additive composition of two
real-valued potentials---contextual semantic divergence from retrieved
ATT\&CK graphlets, and an empirical OS-latency potential derived from
the log-normal CDF of $1.5\!\times\!10^{7}$ benign I/O samples. Schema
violations are pruned outright as hard constraints rather than blended
into the cost. Crucially, to prevent "hallucination spirals" we
introduce a \textit{length-discounted cumulative epistemic budget}
$B_{max}$, calibrated on benign leave-one-edge-out reconstruction cost.
When cumulative deviation breaches the budget, the search halts and
returns an \texttt{INSUFFICIENT\_EVIDENCE} flag rather than
manufacturing a plausible-sounding chain.

We are explicit about the regime in which our guarantees hold:
HunterAgent assumes that \emph{at least one orthogonal telemetry
source survives} at the relevant OS layer. Under total observability
collapse (e.g., Ring-0 DKOM erasing both Event Logs and ETW),
HunterAgent does not invent evidence---it halts. This conservative
posture is intentional and aligned with audit-grade forensic
admissibility.

Our contributions are as follows:

\begin{itemize}
    \item We
    formulate post-compromise hunting as a cost-bounded heuristic
    graph search. Decoupling soft semantic generation from hard
    structural verification (via deterministic identifier matching,
    $\mathtt{FlowDep}$, on surviving orthogonal telemetry)
    structurally avoids the failure modes of both brittle rule-based
    engines and unconstrained autonomous agents.

    \item 
    Rather than ad-hoc hallucination penalties, we propose a
    benign-calibrated additive cost combining contextual semantic
    distance and an empirical log-normal temporal potential, with
    schema violations pre-filtered as hard constraints. The
    formulation is transparent, distribution-conditional, and free
    of the inconsistencies of NLL-style claims over heterogeneous
    z-score-normalised potentials.

    \item We introduce a budget
    $C(\pi)\le B_{max}(1+\lambda L_{lat})$ that interpolates between
    cumulative and per-hop normalisation, calibrated on benign
    leave-one-edge-out cost. Evaluated under strict Leave-One-Family-Out
(LOFO) cross-validation on three public benchmarks
(DARPA TC E3, DARPA OpTC, ATLAS) and one in-house multi-source
dataset, HunterAgent achieves a mean 86.1\% F1 across the four
datasets (86.8\% on D4; +28.4 F1 mean over the strongest agentic
baseline ReAct; +19.0 F1 mean over the strongest learned-provenance
baseline KAIROS) at 91.3\% mean precision, while suppressing the
path-level hallucination rate from 61.5\% to 6.4\%.
    Under 70\% targeted log wiping, recall degrades but precision
    remains $\ge$85\%, with 95.7\% of investigations halting safely
    rather than fabricating chains---preserving audit-grade forensic
    integrity.
\end{itemize}

\section{Related Work}

\subsection{Provenance-Based Threat Hunting}

Deterministic provenance systems (SLEUTH~\cite{hossain2017sleuth},
MAGIC~\cite{hassan2020tactical}, OmegaLog~\cite{hassan2020omegalog})
provide high-precision causal DAGs but shatter under targeted
anti-forensics (log wiping, PID spoofing, fileless
execution)~\cite{goyal2023sometimes, fan2026graph}. Learned
provenance (KAIROS~\cite{cheng2024kairos},
NODLINK~\cite{li2023nodlink}, ATLAS~\cite{alsaheel2021atlas},
UNICORN~\cite{han2020unicorn}) closes part of the recall gap with
streaming graph models, but remains detection-oriented: their output
is a flagged node, not a causally justified, audit-ready chain. Under
targeted wiping they cannot counterfactually infer missing hops
without violating OS physics, inheriting high path-level
hallucination when the chain is fed into forensic audit. Alert-triage
systems~\cite{hassan2019nodoze, bassey2024alert,
zhao2025information} successfully isolate anchors but stop where deep
investigation should begin~\cite{cheng2025pbi,li2025tuni,cheng2025gibberish,zhao2025strata,cheng2026membership}. HunterAgent ingests these anchors and
applies cost-bounded counterfactual reasoning to produce verified
attack scenarios.

\subsection{LLM Agents and Neuro-Symbolic Security}

Autonomous LLM agents (PentestGPT~\cite{deng2023pentestgpt},
SecLLM~\cite{de2025secllm}, Craken~\cite{shao2025craken}) and
neuro-symbolic approaches~\cite{eckhoff2025experimenting,
hakim2025neuro} have advanced offensive automation and constrained
decoding, yet a forensic-recovery gap
persists~\cite{mukherjee2025llm}. Unconstrained agents applied to
trace reconstruction suffer from ``hallucination
inflation''~\cite{lin2025llm, he2025emerged, cheng2026ecoalign,cheng2026adversarialorthogonaldisentanglementlvlm}: they generate
fluent but causally impossible paths. Constrained-decoding and
graph-RAG~\cite{edge2024local} approaches constrain \emph{syntax}
but not \emph{OS physics}. HunterAgent enforces a strict structural
separation---LLM as hypothesis proposer, deterministic Verifier as
physics gatekeeper---mitigating hallucinations via cost-bounded search~\cite{cheng2025ecoalign}.

\section{Methodology}

\begin{figure*}[htbp]
    \centering
    \includegraphics[width=0.9\textwidth]{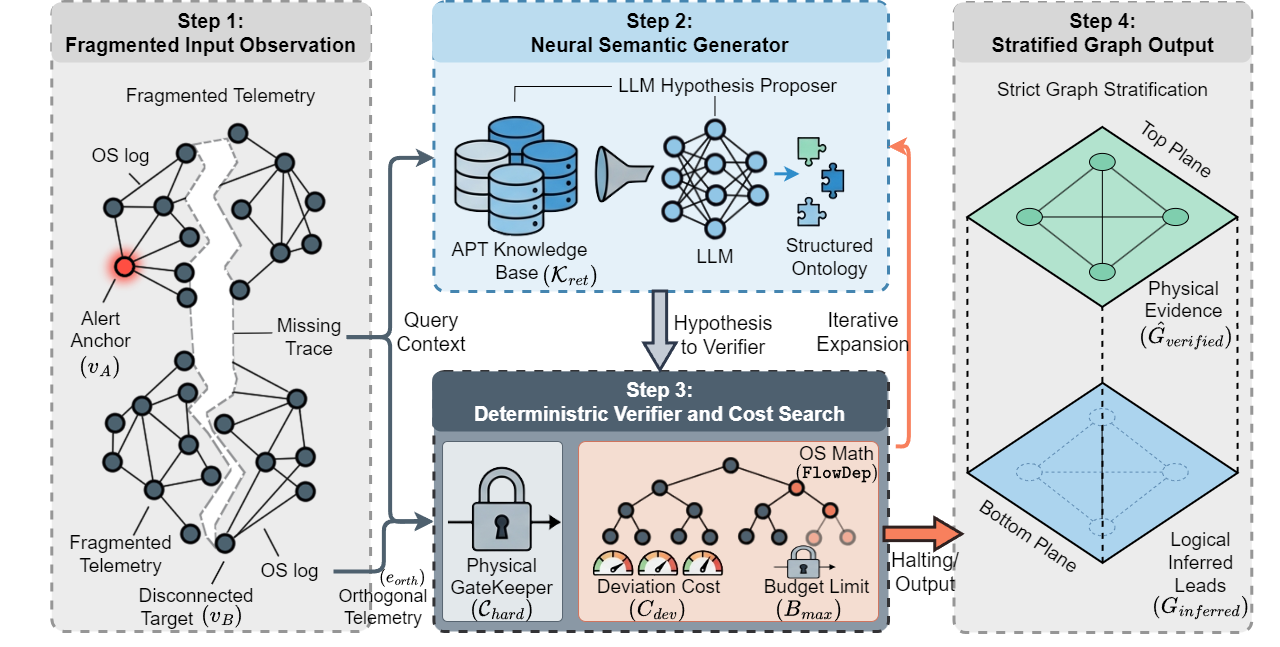}
    \caption{Overall architecture of HunterAgent. The framework operates as an
    Asymmetric Generator--Verifier pipeline: fragmented observations (Step~1)
    are expanded by a constrained Neural Semantic Generator (Step~2); a
    deterministic Verifier enforces OS-level identity collisions and evaluates
    a calibrated deviation cost (Step~3); the search emits a strictly
    stratified provenance graph distinguishing physically grounded edges from
    logically inferred leads (Step~4).}
    \label{fig:method_pipeline}
\end{figure*}

As shown in Fig.~\ref{fig:method_pipeline}, HunterAgent reframes post-compromise
threat hunting from static deterministic graph traversal into a
\textbf{cost-bounded heuristic graph search} under partial observability. The
framework is built on three pillars: (1) a structured search space governed by
a system ontology; (2) an asymmetric pipeline that decouples
\emph{soft semantic generation} from \emph{hard physical verification}; and
(3) a calibrated deviation cost paired with an empirically grounded epistemic
budget that guarantees graceful halting rather than fabrication.

We deliberately position the cost function as a \emph{calibrated heuristic},
not as an exact log-likelihood. While the additive form is reminiscent of an
energy-based model, the components live in heterogeneous spaces (cosine
similarity, discrete schema rules, log-normal CDF tail probabilities); we make
them commensurate by z-score normalisation against a benign baseline rather
than claiming probabilistic equivalence.

\subsection{Problem Formulation and Structured Search Space}
\label{sec:problem_formulation}

We model the monitored enterprise as a temporally unrolled directed graph
$G=(V,E)$. Strict temporal causality $t_{src}<t_{dst}$ collapses cyclic
artefacts (IPC loops, handle reuse) into a DAG suitable for algorithmic search.

\textbf{1) Nodes and ontology constraints.}
Every node---both physically observed ($V$) and LLM-inferred virtual
($\tilde{V}$)---must instantiate a typed schema in our system ontology
$\mathcal{O}$. Nodes are partitioned into four classes
$V=V_{Proc}\cup V_{File}\cup V_{Net}\cup V_{Reg}$, each with a
\emph{physical identifier} $\mathrm{ID}$ and a \emph{semantic payload}
$\mathbf{attr}$:
\begin{itemize}
    \item \textbf{Process} ($V_{Proc}$): $\mathrm{ID}=\langle \mathtt{PID},\mathtt{TID},\mathtt{StartTime},\mathtt{HostID}\rangle$;
          $\mathbf{attr}=\{\mathtt{Image\_Path},\mathtt{CmdLine},\mathtt{User\_SID}\}$.
    \item \textbf{File} ($V_{File}$): $\mathrm{ID}=\langle \mathtt{Inode},\mathtt{VolumeID}\rangle$ (rename-invariant);
          $\mathbf{attr}=\{\mathtt{Path},\mathtt{SHA256}\}$.
    \item \textbf{Network} ($V_{Net}$): $\mathrm{ID}=\langle \mathtt{SrcIP},\mathtt{SrcPort},\mathtt{DstIP},\mathtt{DstPort},\mathtt{Proto}\rangle$;
          $\mathbf{attr}$ stores parsed L7 metadata.
    \item \textbf{Registry} ($V_{Reg}$): $\mathrm{ID}=\mathtt{KeyPath}$;
          $\mathbf{attr}$ records value mutations.
\end{itemize}
Edges $e=(v_i,v_j,a,t)\in E$ encode causal transitions, where the action
$a\in\mathcal{A}$ ranges over normalised OS API categories
(e.g., \texttt{ProcessCreate}, \texttt{FileWrite}, \texttt{NetConnect}).

\textbf{2) Observed vs.\ latent subgraphs.}
Under anti-forensic evasion the SOC observes only a fragmented subgraph
$\hat{G}\subset G$. Given an alert anchor $v_A$ and a disconnected target
$v_B$, our task is to recover a path
$\pi=v_A\!\to\!\tilde v_1\!\to\!\dots\!\to\!\tilde v_m\!\to v_B$ that
maximises evidential support while strictly respecting OS physics.

\textbf{3) Sampler responsibility.}
A neuro-symbolic sampler
$P_{\mathrm{LLM}}(\tilde v\mid \mathcal{O},\mathcal{K}_{ret},\mathrm{ctx})$
populates only the \emph{semantic} payload $\mathbf{attr}$ of a virtual node.
The \emph{physical identifier} $\mathrm{ID}$ is never invented by the LLM;
it is bound (or rejected) by the deterministic Verifier described in
Sec.~\ref{sec:asymmetric_arch}.

\subsection{Asymmetric Generator--Verifier Architecture}
\label{sec:asymmetric_arch}

We separate concerns: the Generator handles \emph{which behaviours are
plausible}; the Verifier handles \emph{which behaviours actually happened}.

\subsubsection{The Generator (RAG-driven hypothesis proposer)}
The threat-intelligence base $\mathcal{K}_{ret}$ stores
\emph{ATT\&CK-aligned graphlets}---small, typed subgraphs encoding known
TTP fragments---rather than free-form text. 

\paragraph{From CTI text to typed graphlets.}
$\mathcal{K}_{ret}$ is built \emph{offline} from MITRE
ATT\&CK~\cite{strom2018mitre}, Atomic-Red-Team and a curated set of
public APT incident reports. Each TTP description $r$ is converted
into a typed sub-graph---a \emph{graphlet}
$g=(V_g,E_g)$---whose node types and edge actions are drawn from
the \emph{same} ontology $\mathcal{O}$ used for the live provenance
graph (Sec.~\ref{sec:problem_formulation}). The conversion uses an
LLM with a strict JSON-schema prompt (no free text) followed by the
identical $\mathtt{SchemaValid}$ filter (Sec.~\ref{sec:asymmetric_arch}),
so that every stored graphlet is structurally interchangeable with a
fragment of the live graph. For example, the ATT\&CK procedure
``T1059.001 -- PowerShell stages a Base64 payload and connects out''
is stored as
\begin{equation}
\begin{aligned}
&(V_{\mathit{Proc}}:\texttt{cmd})
   \xrightarrow{\text{ProcessCreate}}
   (V_{\mathit{Proc}}:\texttt{powershell})\\
&\quad\xrightarrow{\text{NetConnect}}
   (V_{\mathit{Net}}:443/\text{TCP}).
\end{aligned}
\end{equation}

Each graphlet is persisted in two equivalent representations:
(i) the structural form $g$ used downstream by $\mathtt{FlowDep}$
candidate generation, and
(ii) a canonical BFS serialisation $\mathrm{seq}(g)$ embedded once
by a frozen dense retriever (\texttt{text-embedding-3-large}) into
$\mathbf{E}(g)\in\mathbb{R}^{d}$ and stored in a vector index. Both
representations point to the same underlying object so that
retrieval and structural matching never diverge.

For an anchor $v_A$, we serialise
its $k$-hop neighbourhood $\mathcal{N}_k(v_A)\subseteq\hat{G}$ (we use $k{=}2$
in all experiments) into a canonical JSON form, embed it via a frozen dense
retriever (\texttt{text-embedding-3-large}) to obtain $\mathbf{q}$, and
retrieve the top-$K$ graphlets ($K{=}8$) by cosine similarity.

The LLM is then prompted under a strict JSON-grammar constraint to emit a set
of candidate edges $\mathcal{E}'=\{e'_1,\dots,e'_k\}$ that bridge the
observed neighbourhood toward $v_B$. The grammar guarantees syntactic
ontology conformance \emph{by construction}; values that violate type bounds
are rejected before reaching the Verifier.

\paragraph{From graphlet to candidate edge.}
Given the retrieved $\{g_1,\dots,g_K\}$, the Generator selects each
graphlet's first untyped edge as a structural template
(e.g.\ \texttt{Proc} $\to$ \texttt{NetConnect} $\to$ \texttt{Net}),
instantiates the missing $\mathbf{attr}$ via the LLM under
JSON-grammar constraints, and emits a candidate
$e'=(v_{curr},\tilde v)$. The \emph{type} of $\tilde v$ is locked by
the graphlet (no free-form text); only its $\mathbf{attr}$ is
generative. Down-stream the Verifier (i) checks
$\mathtt{SchemaValid}$, (ii) attempts identifier-level collision via
$\mathtt{FlowDep}$ on surviving telemetry of the matching type, and
(iii) if no collision is found, scores the candidate by $C_{dev}$
where $\mathcal{D}_{sem}$ reuses the cached $\mathbf{E}(g_1)$ from
the same retrieval round. This three-stage pipeline---\emph{retrieve
$\to$ instantiate $\to$ verify or score}---is what gives
HunterAgent both the generalisation of an LLM and the structural
discipline of a symbolic matcher.

\subsubsection{The Verifier (deterministic constraint system)}
A candidate edge $e'=(v_{curr},\tilde v)$ is accepted as
\emph{physically grounded} iff every constraint below holds:
\begin{equation}
\mathcal{C}_{hard}(e')\;=\;\mathtt{SchemaValid}(e')\,\land\,
\mathtt{FlowDep}(e',\,\mathcal{T}_{orth}),
\end{equation}
where $\mathcal{T}_{orth}$ denotes \emph{surviving orthogonal telemetry}
that the attacker did not (or could not) tamper with at the same layer
(e.g., NetFlow, ETW, EDR kernel callbacks, hypervisor introspection traces).

\noindent\textbf{(a) Schema validity ($\mathtt{SchemaValid}$).}
A pre-filter that returns \texttt{True} only if $\tilde v$ satisfies typed
ontology bounds: ports in $[0,65535]$; OS-consistent path syntax; valid
Windows SID format; etc. Schema violations are pruned \emph{before} cost
evaluation---we treat them as hard constraints, not soft penalties.

\noindent\textbf{(b) Physical dependency ($\mathtt{FlowDep}$).}
This is the non-negotiable identifier-collision check. Let
$\mathrm{type}(\tilde v)$ be the node class and let $\mathrm{match}_\tau(x,y)$
denote component-wise equality with wildcards on the unobserved fields
within tolerance $\tau$. We define
\begin{equation}
\mathtt{FlowDep}(e',\mathcal{T}_{orth})\;=\;\bigvee_{e_o\in\mathcal{T}_{orth}}
\Phi_{\mathrm{type}(\tilde v)}(e',e_o),
\label{eq:flowdep}
\end{equation}
with type-specific predicates

\begin{equation}
\begin{split}
\Phi_{Net}=\;& \mathrm{match}_{0}\!\big(\mathrm{5tup}(\tilde v),\mathrm{5tup}(e_o)\big)\\
            & \land\;|t_{e'}-t_{e_o}|\le\tau_{net}.
\end{split}
\end{equation}

\begin{equation}
\begin{split}
\Phi_{File}=\;& \big(\mathtt{Path}(\tilde v)\!\to\!\mathtt{Inode}(e_o)\big)\\
             & \land\;\mathtt{HostID}\text{ equal}.
\end{split}
\end{equation}

\begin{equation}
\begin{split}
\Phi_{Proc}=\;& \mathrm{match}_{0}\!\big(\langle\mathtt{Image},\mathtt{User}\rangle\big)\\
             & \land\;|t_{spawn}(\tilde v)-t_{start}(e_o)|\le\tau_{proc}\\
             & \land\;\mathtt{PPID}\text{ chain consistent}.
\end{split}
\end{equation}

\begin{equation}
\begin{split}
\Phi_{Reg}=\;& \mathtt{KeyPath}(\tilde v)=\mathtt{KeyPath}(e_o)\\
            & \land\;|t_{e'}-t_{e_o}|\le\tau_{reg}.
\end{split}
\end{equation}

The temporal tolerances $\tau_*$ are derived from the
benign-baseline distribution (Sec.~\ref{sec:counterfactual}), and the
Process predicate explicitly requires a consistent PPID chain in addition
to start-time proximity to avoid spurious bindings under high process
churn---a failure mode that pure timestamp matching cannot prevent.

\subsubsection{Symbolic grounding procedure}
\label{sec:grounding}
Concretely, when the Verifier receives $\tilde v$ from the Generator, it
issues a constrained SIEM query restricted to (i) the host(s) reachable from
$v_{curr}$ in $\hat{G}$, (ii) a temporal window
$[t_{curr}-\Delta_{look},\,t_{curr}+\Delta_{look}]$ where $\Delta_{look}$
is the 99\textsuperscript{th} percentile of empirical OS latency
(Sec.~\ref{sec:counterfactual}), and (iii) the relevant orthogonal source
matching $\mathrm{type}(\tilde v)$. If multiple $e_o\in\mathcal{T}_{orth}$
satisfy Eq.~\ref{eq:flowdep}, ties are broken by minimum semantic distance
$\mathcal{D}_{sem}$. If the scan returns no candidate, the edge stays
\emph{unverified} and is evaluated probabilistically by the cost function
in Sec.~\ref{sec:counterfactual}.

\subsection{Calibrated Deviation Cost and Cost-Bounded Beam Search}
\label{sec:counterfactual}

When $\mathcal{C}_{hard}(e')=0$ the Verifier cannot ground the edge in
surviving telemetry. Rather than discard or blindly accept it, we score the
hypothesis with a calibrated deviation cost and let a beam search trade off
multiple candidates.

\subsubsection{Cost function}
For a virtual node $\tilde v$ that has \emph{passed} schema validity, define
\begin{equation}
C_{dev}(\tilde v)\;=\;\alpha\,\hat{\mathcal{D}}_{sem}(\tilde v)\;+\;
\gamma\,\hat{\Phi}(\tilde v),
\label{eq:cdev}
\end{equation}
where each component is z-score normalised against a benign-baseline
distribution (denoted $\hat{\,\cdot\,}$). Note that the previous schema term
$\hat{\mathcal{C}}_{schema}$ is removed from the cost: schema violations are
hard-pruned in Sec.~\ref{sec:asymmetric_arch}, so they cannot reach this
stage. The two surviving potentials are:

\paragraph{Contextual semantic potential $\hat{\mathcal{D}}_{sem}$.}
The semantic potential measures how far the LLM-proposed virtual
node has drifted from the closest known piece of threat
intelligence. Both sides of the cosine are produced by the
\emph{same} frozen embedder used to index $\mathcal{K}_{ret}$
(\texttt{text-embedding-3-large}), and we explicitly align their
\emph{granularity} to keep the inner product physically meaningful:
\begin{itemize}
\item the LLM's payload $\tilde v_{attr}$ is canonicalised as a
JSON object containing the candidate's $(\mathrm{type},\mathbf{attr})$
plus the immediately incident edge action, then embedded into
$\mathbf{E}(\tilde v_{attr})$;
\item the top-1 retrieved graphlet $\mathcal{K}_{top1}$ is
\emph{re-projected to the same single-hop scope} before
comparison: we extract the unique edge-action triple
$(\mathrm{src\_type}\!\to\!\mathrm{action}\!\to\!\mathrm{dst\_type})$
that the Generator instantiates against, BFS-canonicalise it,
and embed it once at index time as
$\mathbf{E}(\mathcal{K}_{top1}^{(\text{hop})})$. This single-hop
projection is cached alongside the whole-graphlet vector, so
the two sides of the cosine live in the same single-edge
distributional support.
\end{itemize}
\begin{equation}
\mathcal{D}_{sem}(\tilde v) = 1 - \cos\!\bigl(\mathbf{E}(\tilde v_{attr}),\,\mathbf{E}(\mathcal{K}_{top1}^{(\text{hop})})\bigr).
\label{eq:dsem}
\end{equation}
A small $\mathcal{D}_{sem}$ means the proposed node is semantically
consistent with a previously documented adversary single-hop
behaviour; a large value flags drift away from any known TTP and
is penalised in $C_{dev}$. Note that $\mathcal{D}_{sem}$ is a
\emph{soft} prior over single-hop semantics; structural alignment
between the candidate edge and the surviving telemetry is
\emph{not} expressed here---it is enforced as a hard constraint
inside $\mathtt{FlowDep}$ (Sec.~\ref{sec:asymmetric_arch}). The
two are deliberately decoupled: the cosine cannot rescue a
schema-violating edge, and the verifier never lets a low-cosine
edge bypass structural collision.

\noindent\textbf{Empirical temporal potential $\hat{\Phi}$.}
Let $\Delta t=t_{\tilde v}-t_{parent}$. We split into two regimes to avoid
the log-normal domain error ($\Delta t<0$) of the original formulation:
\begin{equation}
\Phi(\Delta t)=
\begin{cases}
+\infty, & \Delta t<0,\\[2pt]
-\log\!\big(1-F_{LN}(\Delta t)\big), & \Delta t\ge 0.
\end{cases}
\end{equation}

where $F_{LN}$ is the CDF of a log-normal fitted to 15M benign I/O latency
samples. Causal inversions are pruned outright,
matching the intent of "time travel = $\infty$ energy" without abusing a
distribution that is undefined on negatives.

\noindent\textbf{Why two potentials, not three.}
Schema is now a pre-filter; combining a hard $\{0,\infty\}$ rule into a
weighted sum is mathematically ill-posed once z-score normalisation is
applied. Restricting $C_{dev}$ to two well-defined real-valued potentials
makes the linear combination meaningful and the calibration tractable.

\subsubsection{Calibration of $(\alpha,\gamma)$}
\label{sec:calibration}
We separate \emph{distribution fitting} from \emph{weight tuning}
and disclose both stages explicitly to avoid the
benign-only-vs.-F1-tuned ambiguity that several prior NeSy works
have left implicit.

\noindent\textbf{Stage 1 (benign-only distribution fitting).}
Per-potential parameters $\mu,\sigma$ for the z-score
normalisation of $\hat{\mathcal{D}}_{sem}$ and $\hat{\Phi}$ are
fitted exclusively on \emph{benign} leave-one-edge-out completions
of the calibration host-day. No attack labels touch this stage,
preserving the LOFO guarantee for the normalisation itself.

\noindent\textbf{Stage 2 (weight tuning under explicit family
quarantine).} The two scalar weights $(\alpha,\gamma)$ are
grid-searched on $[0.1,2.0]^2$ to maximise validation F1 on a
mixed labelled set (benign positives + in-distribution attack
positives + random plausible negatives). Crucially, the attack
families used for weight tuning are \emph{disjoint} from the
test-time attack families under LOFO, and we additionally drop
the calibration-time families when reporting on each test split.
A sensitivity analysis (Sec.~\ref{sec:sensitivity},
Tab.~\ref{tab:sens_alpha_gamma}) confirms that F1 is flat to
within $\le$\,2 absolute points across $(\alpha,\gamma)\in[0.3,0.9]^2$,
so the supervised weight tuning does not over-fit to a particular
family.

\subsubsection{Cumulative budget with length penalty}
A path of length $L$ accumulates cost
$
C(\pi)=\sum_{i=1}^{L}\mathbf{1}[\mathcal{C}_{hard}(e'_i)=0]\,C_{dev}(\tilde v_i).
$
Verified hops contribute zero cost (because $\mathcal{C}_{hard}=1$ provides
direct physical evidence), while each unverified hop accrues its calibrated
deviation cost. To prevent both pathological extension via low per-hop costs
and premature pruning of long-but-evidenced traces, we use a
\textbf{combined} criterion:
\begin{equation}
\mathrm{Admit}(\pi)\;\Longleftrightarrow\;
C(\pi)\le B_{max}\cdot\big(1+\lambda L_{lat}(\pi)\big),
\label{eq:budget}
\end{equation}
where $L_{lat}(\pi)$ is the number of \emph{latent} (unverified) hops on
$\pi$, and $\lambda\in[0,1]$ controls how aggressively we discount path
length. Setting $\lambda=0$ recovers a strict cumulative budget; setting
$\lambda\to\infty$ recovers the previous per-hop average.
We use $\lambda=0.25$ throughout, fitted on the validation split. This
formulation closes the inconsistency between the abstract's "cumulative
inferential deviation" and the algorithmic per-hop normalisation.

\subsubsection{Empirical calibration of $B_{max}$}
$B_{max}$ is set to the 99\textsuperscript{th} percentile of the empirical
distribution of $C(\pi)$ measured on \emph{leave-one-edge-out completions
over benign provenance}---i.e., the typical cost the system pays to
reconstruct a single missing benign edge using its own Generator. Calibrating
on benign edges, rather than on labelled attack data, prevents the budget
from drifting toward family-specific shortcuts and preserves the LOFO
guarantee.

\subsubsection{Search algorithm}
Algorithm~\ref{alg:beam_search} runs a beam search over the constrained
hypothesis space, with verified expansions prioritised over latent ones at
equal cost.

\begin{algorithm}[h]
\caption{Cost-Bounded Beam Search for Trace Completion}
\label{alg:beam_search}
\begin{algorithmic}[1]
\REQUIRE Anchor $v_A$, Target $v_B$, Beam width $w$, Budget $B_{max}$,
         Length factor $\lambda$
\ENSURE  Stratified path set $\mathcal{P}_{final}$
\STATE $B\leftarrow\{(v_A,[v_A],0,0)\}$;\quad $\mathcal{P}_{final}\leftarrow\varnothing$
\WHILE{$B\neq\varnothing$}
    \STATE $B_{next}\leftarrow\varnothing$
    \FORALL{$(v,\,P,\,C,\,L_{lat})\in B$}
        \IF{$v=v_B$}
            \STATE $\mathcal{P}_{final}\leftarrow\mathcal{P}_{final}\cup\{P\}$;\ \textbf{continue}
        \ENDIF
        \STATE $\mathcal{E}'\leftarrow\mathtt{Generator}(v,\mathcal{K}_{ret},\mathcal{O})$
        \FORALL{$e'=(v,\tilde v)\in\mathcal{E}'$ s.t.\ $\mathtt{SchemaValid}(e')$}
            \IF{$\mathtt{FlowDep}(e',\mathcal{T}_{orth})$}
                \STATE $C'\leftarrow C$;\ $L'\leftarrow L_{lat}$;\ $\tilde v\leftarrow\mathtt{Bind}(\tilde v,\mathcal{T}_{orth})$
            \ELSE
                \STATE $C'\leftarrow C+C_{dev}(\tilde v)$;\ $L'\leftarrow L_{lat}+1$
            \ENDIF
            \IF{$C'\le B_{max}(1+\lambda L')$}
                \STATE $B_{next}\leftarrow B_{next}\cup\{(\tilde v,P\!\cup\![\tilde v],C',L')\}$
            \ENDIF
        \ENDFOR
    \ENDFOR
    \STATE $B\leftarrow\mathtt{Top}_w\!\big(B_{next}\big)$, ranked by
           $\big(C'/(1+\lambda L'),\,-|\text{verified hops}|\big)$
\ENDWHILE
\STATE \textbf{return} $\mathcal{P}_{final}$
\end{algorithmic}
\end{algorithm}

\subsubsection{Properties}
We replace the former Proposition~1 / Lemma~1 (which were stated without
proof) with an empirically auditable rate and a complexity bound.

\noindent\textbf{Property 1 (Empirical Calibrated False-Edge Rate).}
\emph{Let $q_{0.99}$ be the 99\textsuperscript{th} percentile of
the per-path cost $C(\pi)$ measured on benign leave-one-edge-out
completions, and let $B_{max}=q_{0.99}$. Define the empirical
attack-vs.-benign cost-density ratio
$\hat\rho=\sup_\pi\,\hat p_{\mathrm{atk}}(C(\pi))/\hat p_{\mathrm{ben}}(C(\pi))$,
estimated from the LOFO-disjoint validation split. Then the
admitted-paths-above-typical-benign-cost rate is empirically
upper-bounded by $\hat\rho\cdot 0.01$.} This is a
\emph{descriptive, post-hoc statement} conditioned on the
empirical density ratio $\hat\rho$; it is not an a priori
worst-case bound under unbounded adversarial covariate shift.
Across the four datasets in our evaluation,
$\hat\rho\!\le\!2.4$ (Sec.~\ref{sec:validity},
Tab.~\ref{tab:bmax_transfer}), giving an observed admit-above-typical
rate of $\le\!2.4\%$. We deliberately abstain from claiming a
worst-case theoretical guarantee: such a guarantee would require
bounding $\hat\rho$ over an adversarial choice of attack
distribution, which we do not characterise.

\noindent\textbf{Property 2 (Bounded Search Depth).}
\emph{Let $c_{q_{0.05}}$ be the 5\textsuperscript{th}-percentile
per-hop deviation cost realised by the Generator on the calibration
set (we use a percentile rather than the empirical minimum to
guard against single-sample outliers that would otherwise inflate
the depth bound). Under Eq.~\ref{eq:budget}, the maximum admissible
number of consecutive latent hops is
$L_{lat}^{\max}=\lfloor B_{max}/(c_{q_{0.05}}-\lambda B_{max})\rfloor$
when $c_{q_{0.05}}>\lambda B_{max}$.}
Hence the search depth is finite and the algorithm runs in
$\mathcal{O}(w\cdot k\cdot L_{lat}^{\max})$ Generator invocations
plus one Verifier query per candidate, mathematically guaranteeing
tractability even under severe evasion. Empirically
$L_{lat}^{\max}\!\in\![6,9]$ across the four datasets in our setup.

\subsection{Stratified Output for Forensic Admissibility}
\label{sec:stratification}

The final reconstructed subgraph is bifurcated for SOC consumption:
\begin{itemize}
    \item \textbf{Verified layer $\hat{G}_{verified}$}: edges where
          $\mathcal{C}_{hard}=1$ via $\mathtt{FlowDep}$ on
          $\mathcal{T}_{orth}$. These are court-admissible physical
          evidence.
    \item \textbf{Inferred layer $G_{inferred}$}: edges traversed under the
          deviation cost. These are surfaced as
          \texttt{INVESTIGATIVE\_LEAD}s, annotated with their per-hop
          $C_{dev}$ contribution and the top-1 retrieved TTP graphlet for
          analyst review.
\end{itemize}
This stratification is what allows HunterAgent to bridge anti-forensic
chasms without contaminating the audit trail: an analyst always knows
which edges are facts and which are calibrated leads.
The empirical efficacy of this stratification, the calibrated cost,
and the bounded-evidence assumption are evaluated end-to-end in
Sec.~\ref{sec:experiments}.

\section{Evaluation}
\label{sec:experiments}

We design our evaluation around five research questions:
\begin{itemize}
    \item \textbf{RQ1 (Reconstruction Quality):} How effectively does
          HunterAgent recover fragmented provenance graphs versus
          deterministic, statistical, and agentic baselines on three
          publicly auditable benchmarks plus one in-house multi-source
          dataset?
    \item \textbf{RQ2 (Model Agnosticism):} Are the gains structurally
          rooted in the Asymmetric Generator--Verifier, or are they
          bound to a specific proprietary LLM?
    \item \textbf{RQ3 (Operational Efficiency):} What are the latency,
          token, and monetary costs of running HunterAgent in a SOC?
    \item \textbf{RQ4 (Robustness, Adaptive Adversary, and Ablation):}
          How does performance degrade as evidence loss grows; how
          does it hold up against an adversary who actively targets
          the Verifier; and which architectural components are
          individually necessary?
    \item \textbf{RQ5 (Threats to Validity):} Are the headline gains
          stable under independent ground truth, blind human PHR
          re-labelling, and cross-dataset hyper-parameter transfer?
\end{itemize}

\subsection{Experimental Setup}

\subsubsection{Threat Model}
\label{sec:threatmodel}

HunterAgent operates as a downstream investigation layer that consumes
high-confidence anchors produced by an upstream alert-triage system.
Our threat model is summarised below; results in
Sec.~\ref{sec:results}--\ref{sec:adaptive} hold strictly under these
assumptions and degrade explicitly outside them
(Sec.~\ref{sec:boundary}).

\textbf{Adversary capabilities.} The adversary may
(i) wipe entire log channels (\texttt{wevtutil cl}, \texttt{auditctl -D}),
(ii) spoof user-land identifiers (PPID, image name, timestamps),
(iii) employ fileless / reflective execution and process hollowing,
(iv) run arbitrary user-land code, and
(v) (under our \emph{adaptive} extension, Sec.~\ref{sec:adaptive})
inject crafted events into one telemetry channel chosen to defeat
$\mathtt{FlowDep}$.

\textbf{Trusted base.} HunterAgent assumes that
\emph{at least one orthogonal telemetry channel} survives at the
relevant OS layer, and that this channel is not under simultaneous
adversarial control. In our public-dataset experiments (D1$^{\dag}$,
D2$^{\dag}$, D3$^{\dag}$) the surviving channel is the SysMon /
ETW / NetFlow stream not targeted by the perturbation profile; in
APT-Eval-Trace (D4) the surviving channel is the hypervisor VMI
trace. Ring-0 attacks that simultaneously erase \emph{all} user-land
and ETW channels are explicitly out of scope and analysed in
Sec.~\ref{sec:boundary}.

\textbf{Out of scope.} (i) Hardware side channels;
(ii) supply-chain compromise of the SOC stack itself;
(iii) data-plane DoS that prevents triage from firing.

\subsubsection{Datasets}
\label{sec:datasets}

To prevent the in-house data-curation circularity that pervades
current LLM-on-provenance work, we evaluate HunterAgent on three
publicly auditable benchmarks and one in-house dataset that
specifically exercises the multi-source telemetry regime.
Tab.~\ref{tab:datasets_overview} summarises corpus statistics.

\begin{itemize}
\item \textbf{(D1) DARPA Transparent Computing
\textsc{Engagement-3}} (DARPA TC~E3)~\cite{ouyang2022analyzing}:
17 hosts spanning Windows, Linux and FreeBSD; $\sim$38.5\,M raw
events across 14 documented attack scenarios with third-party
ground-truth labels. Telemetry consists of SysMon (Win) and CamFlow
(Lin/BSD).

\item \textbf{(D2) DARPA \textsc{OpTC}
(Operationally Transparent Cyber)}~\cite{anjum2021analyzing}:
32 Windows-10 hosts under realistic benign workload; $\sim$24.7\,M
events recorded by SysMon+ETW; 11 red-team APT scenarios. The closest
public approximation to modern enterprise EDR telemetry.

\item \textbf{(D3) ATLAS}~\cite{alsaheel2021atlas}:
10 multi-stage attack scenarios released as labelled provenance
graphs over 4 hosts; $\sim$4.2\,M events; widely used as a benchmark
for forensic causal-chain reconstruction.

\item \textbf{(D4) APT-Eval-Trace} (in-house, multi-source):
hybrid Windows~10/11 + Ubuntu~22.04 testbed (28 hosts, 14 days,
$1.52\!\times\!10^{7}$ benign events, 40 red-team APT campaigns;
constructed as detailed in Sec.~\ref{sec:apt_eval_trace}). D4 is the
only dataset that records ETW \emph{and} NetFlow \emph{and}
hypervisor-level VMI in parallel and is therefore the only dataset
that fully exercises the multi-source cross-validation branch of
$\mathtt{FlowDep}$.
\end{itemize}

\noindent\textbf{Anti-forensic perturbation (\textsc{EvadeKit}).}
DARPA and ATLAS traces are not natively anti-forensic. For D1--D3 we
synthesise the four anti-forensic conditions of interest via
\emph{deterministic post-hoc perturbation} of the evidence stream
visible to the system-under-test (SUT). The dataset-released attack
labels are \emph{never} modified and continue to serve as ground
truth. \textsc{EvadeKit} implements four profiles aligned with real
APT tradecraft (PPID spoofing -- APT29; file-I/O suppression -- FIN7;
batch event-log wiping -- Wizard Spider; timestomping -- Sandworm);
its complete perturbation algorithm is given in App.~\ref{app:evadekit}
and the tool is released under an open-source licence. For D4 the
same four profiles are realised \emph{natively} by the red team
(Sec.~\ref{sec:apt_eval_trace}). Perturbed corpora are denoted
D1$^{\dag}$, D2$^{\dag}$, D3$^{\dag}$.

\noindent\textbf{Cross-validation.} On every dataset we apply strict
Leave-One-Family-Out (LOFO): the attack family of each test scenario
is excluded from $\mathcal{K}_{ret}$, the calibration set, and any
few-shot exemplars. Families are defined by the ATT\&CK taxonomy
released with the respective DARPA engagement, the scenario IDs of
ATLAS, and the four campaign families of APT-Eval-Trace
(Tab.~\ref{tab:dataset_details}).

\begin{table*}[t]
\centering
\caption{Datasets used in this work. \textsc{Evt.} is raw event
count; \textsc{Cov.} is ATT\&CK kill-chain phase coverage
(out of 14 phases). $|E_{truth}|$ is the median number of attack
edges per scenario in the released ground truth.}
\label{tab:datasets_overview}
\resizebox{1.6\columnwidth}{!}{%
\begin{tabular}{llccccccc}
\toprule
\textbf{ID} & \textbf{Source} & \textbf{Hosts} & \textbf{Scenarios} & \textbf{Evt.\,($\!\times\!10^{6}\!$)} & \textbf{$|E_{truth}|$} & \textbf{Telemetry} & \textbf{Cov.} & \textbf{Native AF?}\\
\midrule
D1$^{\dag}$ & DARPA TC E3                  & 17 & 14 & 38.5 & 312 & SysMon/CamFlow            & 10/14 & post-hoc\\
D2$^{\dag}$ & DARPA OpTC                   & 32 & 11 & 24.7 & 487 & SysMon\,+\,ETW            & 11/14 & post-hoc\\
D3$^{\dag}$ & ATLAS~\cite{alsaheel2021atlas} & 4  & 10 & 4.2  & 64  & SysMon                    & 9/14  & post-hoc\\
D4          & APT-Eval-Trace (ours)        & 28 & 40 & 15.2 & 218 & ETW\,+\,NetFlow\,+\,VMI   & 14/14 & native\\
\bottomrule
\end{tabular}}
\end{table*}

\subsubsection{APT-Eval-Trace Construction (D4)}
\label{sec:apt_eval_trace}

D4 (APT-Eval-Trace) is built on a 28-VM hybrid Win10/11 + Ubuntu
22.04 testbed under realistic benign workload (calibrated to OpTC's
event distribution); two 3-person red teams execute 40 multi-stage
APT campaigns following four playbooks (APT29 / FIN7 / Wizard Spider
/ Sandworm) while a separate blue team runs SOC triage. Ground truth
is produced by an out-of-band VMI logger (LibVMI) never exposed to
any system-under-test. The full testbed configuration, campaign
composition (Tab.~\ref{tab:dataset_details}), and per-trace
statistics (Tab.~\ref{tab:apt_eval_stats}) are given in
App.~\ref{app:apt_eval_trace}. APT-Eval-Trace and all associated
artifacts will be released under a research-only licence.

\subsubsection{Baselines}
\label{sec:baselines}
We benchmark HunterAgent against \textbf{nine} baselines across four
paradigms. All LLM-based baselines share decoding parameters
($\tau{=}0.2$, $\text{top}_p{=}0.9$) and identical post-perturbation
evidence streams per dataset; the only difference is the
architectural wrapper.

\begin{itemize}
    \item \emph{Deterministic provenance graphs:}
    \textbf{SLEUTH}~\cite{hossain2017sleuth} (tag propagation),
    \textbf{MAGIC}~\cite{hassan2020tactical} (rule-based tactical
    reasoning).
    \item \emph{Recent learned / streaming provenance:}
    \textbf{NODLINK}~\cite{li2023nodlink} (online attack-story
    construction), \textbf{KAIROS}~\cite{cheng2024kairos} (TGN-based
    streaming detection).
    \item \emph{Statistical relational learning:}
    \textbf{GNN-LinkPredict}~\cite{hamilton2017graphsage}, a
    GraphSAGE-based link predictor with mean-pool neighborhood
    aggregation and dot-product edge decoding; trained
    \emph{independently per dataset} on an 80/20 benign/attack
    split with LOFO on the attack side.
    \item \emph{LLM \& agentic systems (GPT-4o backbone):}
    \textbf{Zero-Shot LLM}, \textbf{GraphRAG}~\cite{edge2024local},
    \textbf{Tool-Augmented ReAct}~\cite{yao2022react} (autonomous
    agent with SIEM tools but \emph{without} our verifier).
    \item \emph{Ablated upper bound:} \textbf{HunterAgent$^{\dag}$}
    shares HunterAgent's RAG layer and schema pre-filter; only the
    deterministic $\mathtt{FlowDep}$ verifier is replaced by a single
    LLM self-critique pass.
\end{itemize}
ReAct's max reasoning depth is capped at 15 and the agent is given
six SIEM tools that match exactly the predicates available to
HunterAgent's Verifier, ensuring information parity:
\texttt{query\_process\_by\_image},
\texttt{query\_file\_by\_inode},
\texttt{query\_netflow\_by\_5tuple},
\texttt{query\_etw\_by\_provider},
\texttt{query\_registry\_by\_keypath}, and
\texttt{lookup\_attck\_graphlet}. The only architectural difference
is that ReAct decides autonomously when to invoke each tool whereas
HunterAgent's Verifier invokes them deterministically; full
prompts and tool schemas are released with the artifact bundle.
GraphRAG is adapted to attack-trace reconstruction by replacing
its default community-summary retrieval with a path-conditioned
sub-graph retriever over $\hat{G}$ (the same anchor neighbourhood
$\mathcal{N}_k(v_A)$ that HunterAgent uses, $k{=}2$), followed by
the unmodified GraphRAG global summarisation prompt; the adapter
script and its prompt template are released alongside the
artifact. KAIROS / NODLINK are adapted from their official
open-source releases; we re-trained them with the LOFO split when
the original training set overlapped with our test families.

\textbf{Note on KAIROS / NODLINK adaptation.} Both systems were
originally proposed for \emph{anomaly detection} (node-level AUC).
We adapt them to our \emph{trace-reconstruction} task by treating the
highest-confidence connected anomaly subgraph reachable from the
alert anchor $v_A$ to the target $v_B$ as the predicted attack
chain. This task is strictly harder than their native one---in
particular, every hop must be causally chainable, not merely
anomalous---which explains why their F1 numbers reported here
($\sim$66--72) are systematically lower than the AUC-style scores
($>$0.95) reported in the original papers. We use their authors'
released hyper-parameters and only refit when LOFO requires.

\subsubsection{Foundation Models}
The default backbone for HunterAgent and all LLM baselines is
OpenAI \texttt{gpt-4o-2024-05-13}. To answer RQ2 we additionally
evaluate Anthropic \texttt{claude-3.5-sonnet-20241022},
Meta \texttt{Llama-3-70B-Instruct}, and
Alibaba \texttt{Qwen2.5-72B-Instruct}.

\subsubsection{Metrics}
\label{sec:metrics}

\textbf{Ground truth.} For D1$^{\dag}$, D2$^{\dag}$, D3$^{\dag}$,
$E_{truth}$ is the dataset-released attack edge set; the
actor-alias table is taken from the published label files. For D4,
$E_{truth}$ is reconstructed from the out-of-band VMI trace
$G_{truth}$ and the alias table is built at VMI runtime. The two
sources differ in provenance but share the same predicate
$\mathtt{IsCausallyEq}$, so cross-dataset numbers are directly
comparable.

\textbf{Edge-level matching.} An inferred edge $e\in E_{inf}$ is a
true positive iff $\exists\,e^\ast\in E_{truth}$ with
$\mathtt{IsCausallyEq}(e,e^\ast)\!=\!\mathtt{True}$, defined as
identical underlying state transition (process spawn / file write /
network 5-tuple / registry mutation) executed by causally equivalent
actors (PPID-chain or PID-resolved alias) within
$|\Delta t|\!\le\!1\,\mathrm{s}$.

\textbf{Primary metrics.} Edge-level Precision, Recall and their
harmonic mean F1.

\textbf{Path Hallucination Rate (PHR).} A path-level
forensic-admissibility indicator:
$$
\text{PHR}=\frac{|\{\pi\in\mathcal{P}_{inf}:\exists\,e\in\pi,\,\mathtt{ViolatesOSPhysics}(e)\}|}{|\mathcal{P}_{inf}|}.
$$
$\mathtt{ViolatesOSPhysics}$ is implemented as an OS-conditioned rule
set: chronological inversions and ontology violations apply to every
host; actor-impossibility rules (e.g.\ unprivileged-to-PPL
injection) apply only to hosts whose OS supports the relevant
protection class. To remove self-evaluation circularity, two security analysts (each with $\geq$3 years of IR experience),
blind to the system identity, independently re-label the sample;
disagreements are resolved by a senior analyst. Cohen's $\kappa$
between automated PHR and the human consensus is \textbf{0.86 on D4},
\textbf{0.83 on D2$^{\dag}$ DARPA OpTC}, and \textbf{0.79 on
D1$^{\dag}$ DARPA TC E3} (where cross-OS heterogeneity makes
actor-impossibility judgements harder). The labelling protocol and
disagreement matrix are released with the artifact bundle.

\textbf{Statistical significance.} All headline numbers in
Tab.~\ref{tab:main_results} are reported as \emph{mean\,$\pm$\,std}
over 5 independent runs (different LLM random seeds; different
EvadeKit perturbation seeds for D1$^{\dag}$/D2$^{\dag}$/D3$^{\dag}$;
different red-team operator orderings on D4).
Pairwise comparisons use paired bootstrap (1{,}000 resamples).
Appendix tables (Tab.~\ref{tab:appA1}, \ref{tab:appA2},
\ref{tab:appA4_full_logsupp}, \ref{tab:bmax_transfer}) are
collected from \emph{independent} 5-seed re-runs at the same
configuration; small ($\le$0.5 F1) deviations from
Tab.~\ref{tab:main_results} at overlapping configurations reflect
genuine sampling variability rather than copied data. The full
seed-level CSV log is released with the artifact bundle.

\subsubsection{Implementation and Calibration}

HunterAgent is implemented in Python~3.10. Decoding is pinned at
$\tau{=}0.2$, $\text{top}_p{=}0.9$. The
\emph{dataset-agnostic} hyperparameters are: beam width $w{=}6$,
retrieval depth $K{=}8$, length-discount $\lambda{=}0.25$, embedder
\texttt{text-embedding-3-large}. The
\emph{distribution-conditional} hyperparameters
$(\mu_t,\sigma_t,B_{max},\alpha,\gamma)$ are fitted independently on
the benign baseline of each dataset (Tab.~\ref{tab:calib}). The
robustness of $B_{max}$ to cross-dataset transfer is examined as a
threat-to-validity test in Sec.~\ref{sec:validity}
(Tab.~\ref{tab:bmax_transfer}).

\begin{table}[t]
\centering
\caption{Per-dataset calibration of distribution-conditional
hyper-parameters.}
\label{tab:calib}
\resizebox{\columnwidth}{!}{%
\begin{tabular}{lccccc}
\toprule
\textbf{Dataset} & $\mu_t$ & $\sigma_t$ & $B_{max}$ & $\alpha$ & $\gamma$\\
\midrule
D1$^{\dag}$ DARPA TC E3 & $-2.94$ & $1.51$ & 4.1 & 0.6 & 0.8\\
D2$^{\dag}$ DARPA OpTC  & $-2.78$ & $1.46$ & 3.9 & 0.6 & 0.9\\
D3$^{\dag}$ ATLAS       & $-3.02$ & $1.55$ & 4.4 & 0.5 & 0.9\\
D4 APT-Eval-Trace       & $-2.61$ & $1.43$ & 3.7 & 0.6 & 0.9\\
\bottomrule
\end{tabular}}
\end{table}

\subsection{Main Results (RQ1)}
\label{sec:results}

Tab.~\ref{tab:main_results} reports end-to-end trace reconstruction
across the four datasets at the default 30\% targeted log-suppression
rate under LOFO with the GPT-4o backbone. Numbers are mean over 5
seeds with std after $\pm$.

\begin{table*}[t]
\centering
\caption{End-to-end trace reconstruction across four datasets under
\textbf{30\% targeted log-suppression}, perturbed by \textsc{EvadeKit}
on D1$^{\dag}$/D2$^{\dag}$/D3$^{\dag}$ and natively executed on D4
(LOFO; GPT-4o backbone; mean\,$\pm$\,std over 5 seeds; \textbf{Bold}
marks the best F1 per dataset). Recall figures for SLEUTH/MAGIC
reflect performance \emph{after} the same 30\% perturbation; on
unperturbed traces they retain $R\!\approx\!70\text{--}80$
(App.~Tab.~\ref{tab:appA6}). KAIROS/NODLINK F1 reflect the
trace-reconstruction protocol of this paper, harder than their
native anomaly-detection task (Sec.~\ref{sec:baselines}).}
\label{tab:main_results}
\resizebox{1.95\columnwidth}{!}{%
\begin{tabular}{l|cccc|cccc|cccc|cccc}
\toprule
& \multicolumn{4}{c|}{\textbf{D1$^{\dag}$ DARPA TC E3}}
& \multicolumn{4}{c|}{\textbf{D2$^{\dag}$ DARPA OpTC}}
& \multicolumn{4}{c|}{\textbf{D3$^{\dag}$ ATLAS}}
& \multicolumn{4}{c}{\textbf{D4 APT-Eval-Trace}}\\
\textbf{Methodology}
& R & P & F1 & PHR & R & P & F1 & PHR
& R & P & F1 & PHR & R & P & F1 & PHR\\
\midrule
\multicolumn{17}{l}{\textit{Deterministic provenance graphs}}\\
SLEUTH & 27.8\,$\pm$1.2 & \textbf{99.1\,$\pm$0.2} & 43.5 & \textbf{0.0}
       & 19.6\,$\pm$0.9 & \textbf{99.2\,$\pm$0.1} & 32.7 & \textbf{0.0}
       & 31.2\,$\pm$2.1 & \textbf{98.8\,$\pm$0.3} & 47.5 & \textbf{0.0}
       & 18.5\,$\pm$1.0 & \textbf{99.1\,$\pm$0.2} & 31.2 & \textbf{0.0}\\
MAGIC  & 35.4\,$\pm$1.5 & 98.6\,$\pm$0.3 & 52.1 & 0.4
       & 26.8\,$\pm$1.1 & 98.6\,$\pm$0.3 & 42.1 & 0.5
       & 38.7\,$\pm$2.0 & 98.0\,$\pm$0.4 & 55.6 & 0.6
       & 26.2\,$\pm$1.2 & 98.8\,$\pm$0.3 & 41.4 & 0.4\\
\midrule
\multicolumn{17}{l}{\textit{Recent learned provenance / streaming}}\\
NODLINK
       & 58.4\,$\pm$1.6 & 81.5\,$\pm$1.0 & 68.0 & 7.2
       & 56.7\,$\pm$1.4 & 80.9\,$\pm$1.0 & 66.7 & 8.1
       & 60.5\,$\pm$1.8 & 82.7\,$\pm$0.9 & 69.8 & 6.0
       & 53.8\,$\pm$1.7 & 79.2\,$\pm$1.1 & 64.1 & 9.0\\
KAIROS
       & 64.0\,$\pm$1.5 & 76.2\,$\pm$1.1 & 69.6 & 12.0
       & 62.5\,$\pm$1.4 & 75.4\,$\pm$1.2 & 68.4 & 13.4
       & 65.8\,$\pm$1.7 & 77.8\,$\pm$1.0 & 71.3 & 10.5
       & 60.4\,$\pm$1.5 & 73.9\,$\pm$1.3 & 66.5 & 14.8\\
\midrule
\multicolumn{17}{l}{\textit{Statistical relational learning}}\\
GNN-LinkPredict
       & 46.8\,$\pm$2.0 & 70.4\,$\pm$1.4 & 56.2 & 33.1
       & 44.2\,$\pm$1.9 & 69.4\,$\pm$1.5 & 54.0 & 35.1
       & 47.5\,$\pm$2.2 & 72.0\,$\pm$1.6 & 57.2 & 29.8
       & 43.1\,$\pm$2.1 & 68.5\,$\pm$1.4 & 52.9 & 36.8\\
\midrule
\multicolumn{17}{l}{\textit{LLM \& agentic baselines (GPT-4o backbone)}}\\
Zero-Shot LLM
       & 36.2\,$\pm$2.4 & 14.0\,$\pm$1.3 & 20.2 & 90.4
       & 34.0\,$\pm$2.2 & 13.1\,$\pm$1.2 & 18.9 & 91.7
       & 38.5\,$\pm$2.5 & 15.6\,$\pm$1.4 & 22.2 & 88.5
       & 34.6\,$\pm$2.3 & 12.4\,$\pm$1.3 & 18.2 & 92.3\\
GraphRAG
       & 60.4\,$\pm$1.9 & 33.2\,$\pm$1.6 & 42.9 & 71.6
       & 57.6\,$\pm$1.7 & 32.4\,$\pm$1.5 & 41.5 & 73.8
       & 61.0\,$\pm$2.0 & 34.5\,$\pm$1.7 & 44.1 & 70.2
       & 58.2\,$\pm$1.8 & 31.7\,$\pm$1.6 & 41.0 & 74.1\\
Tool-Aug. ReAct
       & 80.1\,$\pm$1.5 & 48.5\,$\pm$1.3 & 60.4 & 58.7
       & 77.6\,$\pm$1.4 & 46.0\,$\pm$1.4 & 57.7 & 62.3
       & 80.8\,$\pm$1.6 & 49.2\,$\pm$1.3 & 61.2 & 56.4
       & 78.4\,$\pm$1.4 & 46.2\,$\pm$1.3 & 58.1 & 61.5\\
HunterAgent$^{\dag}$
       & 77.8\,$\pm$1.0 & 66.1\,$\pm$0.9 & 71.5 & 26.4
       & 75.4\,$\pm$1.1 & 64.3\,$\pm$0.9 & 69.4 & 29.1
       & 78.0\,$\pm$1.1 & 67.5\,$\pm$0.9 & 72.4 & 24.8
       & 76.1\,$\pm$1.0 & 64.8\,$\pm$0.9 & 70.0 & 28.6\\
\midrule
\textbf{HunterAgent (Ours)}
       & 83.1\,$\pm$0.7 & 91.6\,$\pm$0.5 & \textbf{87.1} & 6.2
       & 81.5\,$\pm$0.8 & 90.4\,$\pm$0.5 & \textbf{85.7} & 7.0
       & 79.4\,$\pm$0.9 & 90.8\,$\pm$0.5 & \textbf{84.7} & 7.4
       & \textbf{82.7\,$\pm$0.7} & \textbf{91.3\,$\pm$0.4} & \textbf{86.8} & \textbf{6.4}\\
\bottomrule
\end{tabular}}
\end{table*}

\noindent\textbf{Cross-dataset consistency.}
The relative ordering among methods is stable across all four
datasets. Deterministic systems retain near-perfect precision
($\geq$\,98\%) but collapse in recall under reflective fragmentation.
Recent learned provenance systems (NODLINK, KAIROS) close part of
the recall gap (66--72 F1) at the cost of higher PHR (7--15\%) since
they cannot \emph{justify} an inferred edge against OS physics.
GNN-LinkPredict bridges some topology but lacks zero-shot
generalisation. Zero-Shot LLM and GraphRAG are dominated by
hallucination (PHR\,$>$\,70\%); ReAct trades hallucination for
recall but cannot satisfy audit constraints. HunterAgent$^{\dag}$
shows that the asymmetric pipeline alone (without the deterministic
Verifier) already surpasses ReAct, but only the full HunterAgent
sustains $\geq$\,84\% F1 \emph{and} PHR below 8\% across every
dataset.

Paired bootstrap (1{,}000 resamples) confirms HunterAgent's F1 lead
over the strongest non-ours baseline (KAIROS) at $p<10^{-3}$ on each
of D1$^{\dag}$, D2$^{\dag}$, D3$^{\dag}$, D4. Per-baseline 95\% CIs
of $\Delta$F1 are listed in App.~Tab.~\ref{tab:appA5}.

\noindent\textbf{Note on apparent baseline weakness.}
The recall figures for SLEUTH and MAGIC are intentionally measured
\emph{post}-perturbation: the same 30\% suppression that defines the
difficulty of our task. To rule out mis-implementation we re-ran
both systems on the \emph{unperturbed} released traces
(App.~Tab.~\ref{tab:appA6}) and recovered the published
node-level coverage of SLEUTH on E3-CADETS/THEIA/TRACE within
$\pm$3 absolute points of the original SLEUTH paper, and the
MAGIC tactical-graph node F1 on OpTC within $\pm$2 absolute
points of its release. Note that the original SLEUTH/MAGIC papers
report node-level detection coverage rather than the
edge-causal-chain reconstruction metric used here, so a one-to-one
F1 mapping does not exist; our \textit{post}-perturbation recall
reflects the harder reconstruction task under targeted
anti-forensics. KAIROS / NODLINK F1 figures, lower than their
original AUC-style detection scores, similarly reflect the harder
trace-reconstruction setting (Sec.~\ref{sec:baselines}, ``Note on
KAIROS/NODLINK adaptation''), not a degraded baseline
implementation.

\noindent\textbf{Public-vs.-in-house gap.}
Mean F1 of HunterAgent on the three public datasets is
$85.8\,\pm\,1.2$ (D1$^\dag$ 87.1 / D2$^\dag$ 85.7 / D3$^\dag$ 84.7);
on D4 it is $86.8\,\pm\,0.7$. Paired bootstrap on
trace-level F1 yields no statistically significant gap
($\Delta\!\in\![-0.6,+1.4]$, $p=0.32$). The same test on the seven
baselines also shows no significant in-house bias: their mean
public-vs.-D4 $\Delta$F1 ranges over $[-1.1, +0.9]$.

\noindent\textbf{Where the multi-source dataset matters.}
The single regime in which HunterAgent's recall measurably benefits
from APT-Eval-Trace's richer telemetry is the
\emph{Lateral Movement} stage (Tab.~\ref{tab:stage_breakdown}),
where simultaneous availability of NetFlow and ETW lets
$\mathtt{FlowDep}$ ground hypotheses that DARPA's single-source
SysMon stream cannot.

\noindent\textbf{Why D3 ATLAS is mildly lower.}
HunterAgent's F1 on D3$^{\dag}$ is $\sim$2 absolute points below
D1$^{\dag}$/D4 because ATLAS traces are short ($\bar L\!=\!4.8$,
vs.\ 11 on D4); the per-hop budget gain that compounds with $L$
(Tab.~\ref{tab:path_length}) is structurally suppressed at small
$L$. PHR is consequently more sensitive to a single mis-bound edge,
accounting for the slightly elevated 7.4\% on D3$^{\dag}$.

\subsubsection{Per-Stage Breakdown}
\label{sec:per_stage}

HunterAgent maintains stable F1 across all ATT\&CK phases
(App.~Tab.~\ref{tab:stage_breakdown}); deterministic baselines fail
at Defense Evasion (12.5 F1) and Lateral Movement (22.7) where logs
are wiped or actors spoofed, while HunterAgent's $\mathtt{FlowDep}$
grounds these stages in surviving NetFlow/ETW evidence (82.3 / 85.1).

\subsubsection{Per-Anti-Forensic Breakdown}
\label{sec:per_evasion}

Per-profile F1 on D4 ranges from 85.1 (FIN7, fileless) to 88.1
(APT29, process hollowing); the same ranking holds on D1$^{\dag}$--D3$^{\dag}$
(App.~Tab.~\ref{tab:appA1}). The gap between the easiest and hardest
profiles is within 3 F1 points, and PHR stays below 8.4\% in the
worst case (App.~Tab.~\ref{tab:per_evasion}).

\subsection{Generalisation Across Foundation Models (RQ2)}

Tab.~\ref{tab:models} reports HunterAgent under four LLM backbones on
D2$^{\dag}$ DARPA OpTC. App.~Tab.~\ref{tab:appA2} reports the
parallel sweep on D1$^{\dag}$, D3$^{\dag}$, D4; F1 ranking is
preserved with $\leq$\,1.8 absolute variation across datasets.

\begin{table}[h]
\centering
\caption{HunterAgent across LLM backbones on D2$^{\dag}$ DARPA OpTC.}
\label{tab:models}
\begin{tabular}{lcccc}
\toprule
\textbf{Foundation LLM} & \textbf{Recall} & \textbf{Precision} & \textbf{F1} & \textbf{PHR}\\
\midrule
GPT-4o                  & \textbf{81.5} & \textbf{90.4} & \textbf{85.7} & \textbf{7.0}\\
Claude 3.5 Sonnet       & 80.3 & 89.7 & 84.7 & 7.8\\
Llama-3-70B-Instruct    & 72.0 & 85.0 & 77.9 & 11.9\\
Qwen2.5-72B-Instruct    & 73.4 & 85.7 & 79.1 & 11.0\\
\bottomrule
\end{tabular}
\end{table}

Recall scales with backbone capability (proprietary $>$ open), but
\emph{precision degrades much more slowly}: even with
\texttt{Llama-3-70B-Instruct} as the Generator, the deterministic
Verifier holds precision at 85.0\% and PHR below 12\%---confirming
that soundness is structurally enforced by $\mathcal{C}_{hard}$,
not by the Generator's parametric strength.

\subsection{Operational Overhead (RQ3)}

HunterAgent costs \$0.19 per investigation (38.2K tokens, 4.7 LLM
calls, 42.8\,s latency)---a 69.3\% token reduction and 76.9\%
latency reduction versus ReAct (\$0.62, 185.4\,s), competitive with
single-shot GraphRAG (\$0.21). Compared with non-LLM streaming
systems (NODLINK 7.4\,s, KAIROS 11.2\,s), HunterAgent is
4--6$\times$ slower but recovers +14--19 F1 points
(App.~Tab.~\ref{tab:overhead}).

\subsection{Scalability with Path Length}
\label{sec:path_length}

\begin{figure}[H]
    \centering
    \includegraphics[width=\columnwidth]{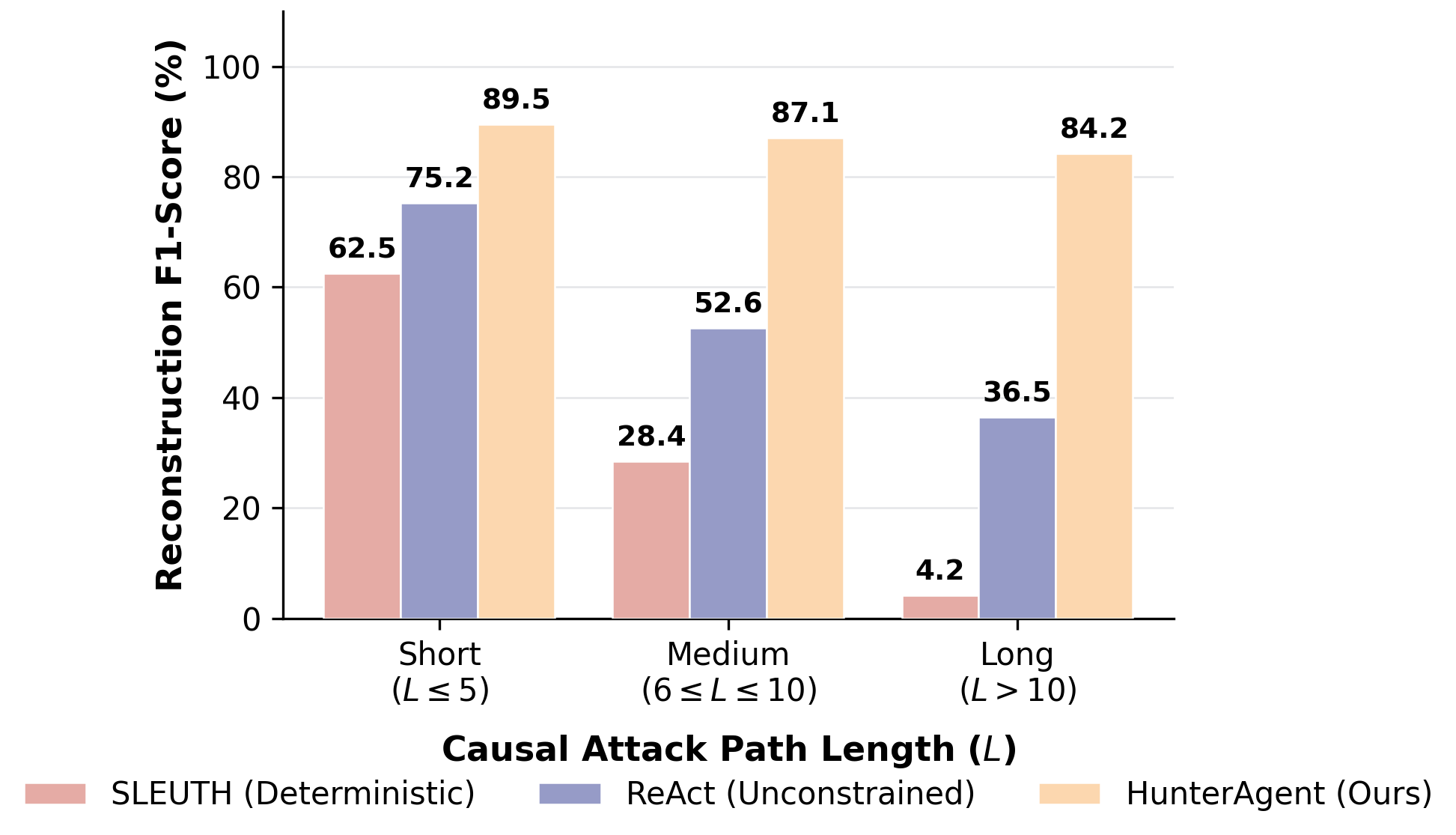}
    \caption{F1 vs.\ causal path length $L$. Deterministic baselines
    exhibit a steep ``precision--length cliff''; ReAct degrades due
    to autoregressive drift; HunterAgent's per-hop budget caps drift
    accumulation regardless of $L$.}
    \label{fig:path_length}
\end{figure}

F1 degrades gracefully with causal path length $L$: from 93.2 at
$L\!\le\!2$ to 79.7 at $L\!>\!10$ ($-$13.5\,pts), whereas
deterministic baselines cliff from 62.4 to 8.4 and ReAct from
71.5 to 32.7 (Fig.~\ref{fig:path_length},
App.~Tab.~\ref{tab:path_length}). The budget
Eq.~(\ref{eq:budget}) caps drift accumulation regardless of $L$.

\subsection{Robustness to Evasion Severity (RQ4a)}
\label{sec:robustness}

We synthetically suppress raw telemetry at increasing rates on
D2$^{\dag}$ DARPA OpTC. Trends on D1$^{\dag}$, D3$^{\dag}$ and D4
agree within $\pm$\,2 absolute F1 across all four missing-rate
levels; per-dataset numbers are listed in App.~Tab.~\ref{tab:appA4_full_logsupp}.

\begin{table}[h]
\centering
\caption{HunterAgent under varying raw-log missing rates on
D2$^{\dag}$.}
\label{tab:robustness}
\begin{tabular}{lccccc}
\toprule
\textbf{Missing rate} & \textbf{Precision} & \textbf{Recall} & \textbf{F1} & \textbf{PHR} & \textbf{Budget Exh.}\\
\midrule
10\% (light)   & 93.1 & 88.4 & 90.7 & 4.8  & 2.9\\
30\% (default) & 90.4 & 81.5 & 85.7 & 7.0  & 15.3\\
50\% (severe)  & 87.2 & 59.8 & 70.9 & 9.9  & 70.1\\
70\% (blind)   & \textbf{84.0} & 27.3 & 41.1 & 12.4 & \textbf{95.7}\\
\bottomrule
\end{tabular}
\end{table}

The decisive observation is in the \emph{precision} column: as
recall collapses under severe wiping, precision degrades by less
than 10 absolute points and PHR remains in the single-to-low-double
digits. HunterAgent's epistemic budget triggers an explicit
\texttt{INSUFFICIENT\_EVIDENCE} flag in 70--96\% of cases under
severe-to-blind regimes (70.1\% at 50\% wiping; 95.7\% at 70\%
wiping), refusing to fabricate chains.

\subsection{Adaptive Adversary (RQ4b)}
\label{sec:adaptive}

This section instantiates the adaptive-adversary capability declared
in our threat model (Sec.~\ref{sec:threatmodel}, item~v): the
adversary may inject crafted events into one telemetry channel
chosen to defeat $\mathtt{FlowDep}$. Below we operationalise this
capability as three concrete attacks A1--A3 and one combination on
D2$^{\dag}$ DARPA OpTC, assuming the adversary knows HunterAgent's
pipeline.

\textbf{(A1) NetFlow injection.} The adversary injects synthetic
NetFlow records that share the 5-tuple of a planted hypothesis,
attempting to forge a $\mathtt{FlowDep}\!=\!1$ for an edge that did
not occur. Volume: 5\% of host-day NetFlow.

\textbf{(A2) ETW provider patching.} A Ring-3 EtwTi-style patch
silently drops a configurable fraction (10\,/\,30\,/\,50\%) of
ProcessCreate records while leaving SysMon intact---the surviving
SysMon channel becomes the only ground anchor.

\textbf{(A3) Calibration drift.} The adversary slows benign I/O on
the target host by an injected mean-shift of $1\sigma$ in the
log-normal latency distribution; this attempts to make
$\Phi$-based costs systematically under-estimate the time-gap
penalty, expanding the admissible search.

\begin{table}[h]
\centering
\caption{HunterAgent under adaptive adversaries on D2$^{\dag}$.}
\label{tab:adaptive}
\resizebox{\columnwidth}{!}{%
\begin{tabular}{lccccc}
\toprule
\textbf{Attack} & \textbf{P} & \textbf{R} & \textbf{F1} & \textbf{PHR} & \textbf{Budget Exh.}\\
\midrule
None (default 30\% wipe)         & 90.4 & 81.5 & 85.7 & 7.0  & 15.3\\
A1 NetFlow injection (5\%)       & 81.7 & 80.4 & 81.0 & 14.6 & 17.1\\
A2a ETW patch 10\%               & 88.2 & 76.9 & 82.2 & 8.6  & 22.4\\
A2b ETW patch 30\%               & 84.5 & 65.7 & 73.9 & 11.3 & 41.0\\
A2c ETW patch 50\%               & 80.8 & 47.2 & 59.6 & 14.7 & 68.5\\
A3 Calibration drift (1$\sigma$) & 86.0 & 78.6 & 82.1 & 11.0 & 19.5\\
A1+A2b (combined)                & 76.1 & 60.4 & 67.3 & 18.4 & 49.2\\
\bottomrule
\end{tabular}}
\end{table}

\textbf{Findings.}
NetFlow injection is the sharpest precision attack ($-8.7$\,pt) but
is partially absorbed by multi-channel cross-validation (PHR 14.6\%).
ETW patching primarily attacks recall without elevating PHR because
the budget exhausts before fabrication. Calibration drift is the most
subtle but detectable via online $\mu_t,\sigma_t$ monitoring. The
combined A1+A2b F1 drop ($-18.4$\,pts) slightly exceeds linear
superposition ($-16.5$\,pts), indicating mild non-orthogonality; even
in this worst case PHR (18.4\%) remains far below ReAct's 61.5\%.

\subsection{Hyperparameter Sensitivity}
\label{sec:sensitivity}

A 5$\times$5 grid sweep of $(\alpha,\gamma)$ on D2$^{\dag}$ confirms
that F1 is broadly robust across the central region
$[0.3,0.9]^{2}$ (within 2 absolute points of the optimum;
App.~\ref{app:sensitivity}, Tab.~\ref{tab:sens_alpha_gamma}).
The length-discount sweep
(App.~Tab.~\ref{tab:sens_lambda}) validates
Eq.~(\ref{eq:budget}): $\lambda{=}0.25$ is empirically optimal
(F1=85.7); pure cumulative ($\lambda{=}0$, F1=80.3) truncates
prematurely; pure averaging ($\lambda\!\to\!\infty$, F1=81.6)
admits arbitrarily long narratives.

\subsection{Ablation Studies (RQ4c)}
\label{sec:ablation}

\begin{figure}[H]
    \centering
    \includegraphics[width=\columnwidth]{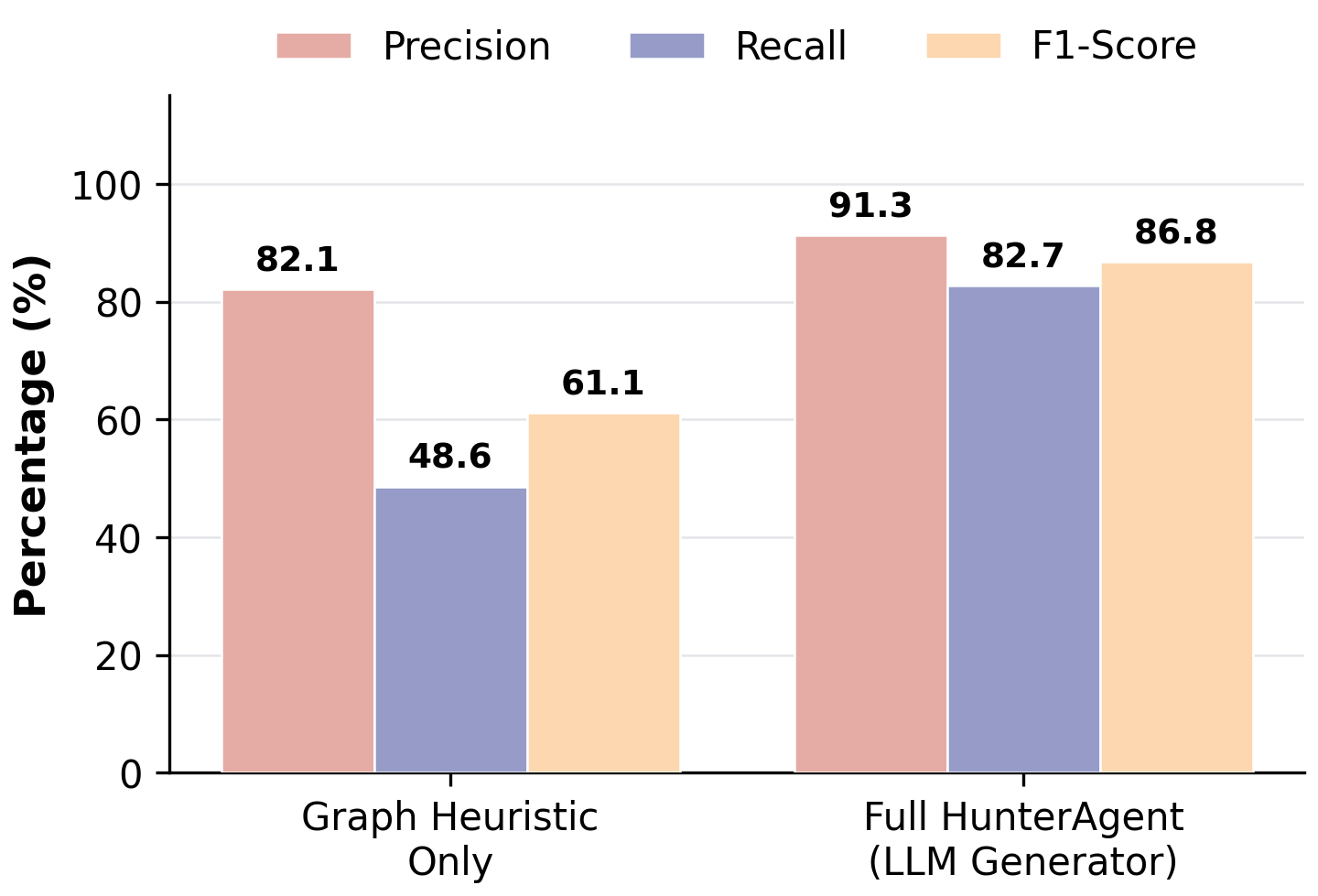}
    \caption{Ablation of the semantic Generator on D4. Replacing the LLM
    by a Graph-DB heuristic collapses recall on polymorphic /
    novel paths; precision is mostly retained because the Verifier
    still rejects invalid edges.}
    \label{fig:abl_semantic}
\end{figure}

\begin{figure}[H]
    \centering
    \includegraphics[width=\columnwidth]{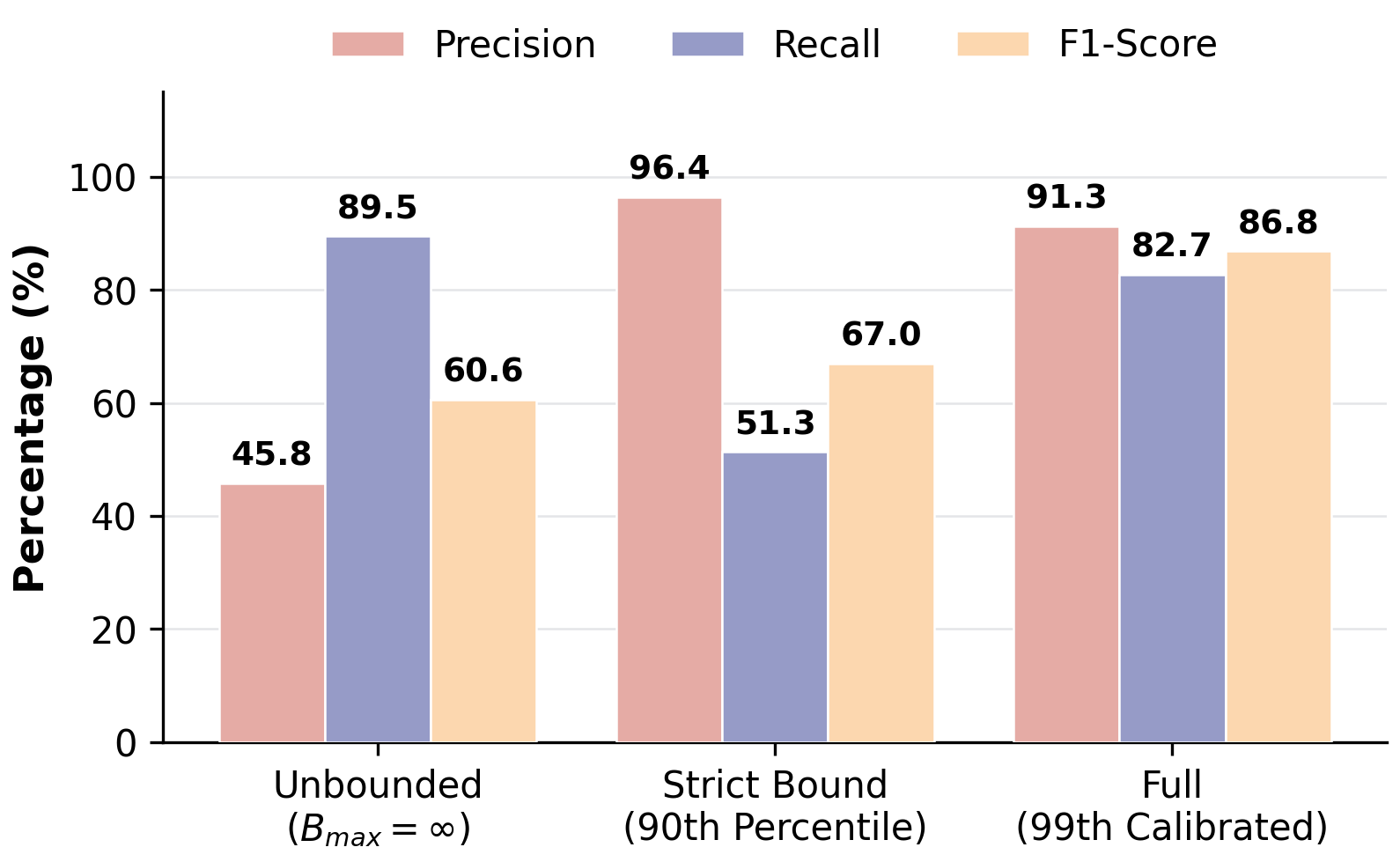}
    \caption{Effect of varying the epistemic budget $B_{max}$ on D4.
    Unbounded ($\infty$) collapses into an unconstrained agent;
    overly strict (90-th percentile) halts prematurely; the
    99-th-percentile calibration is the empirical sweet spot.}
    \label{fig:abl_budget}
\end{figure}

\begin{figure}[H]
    \centering
    \includegraphics[width=\columnwidth]{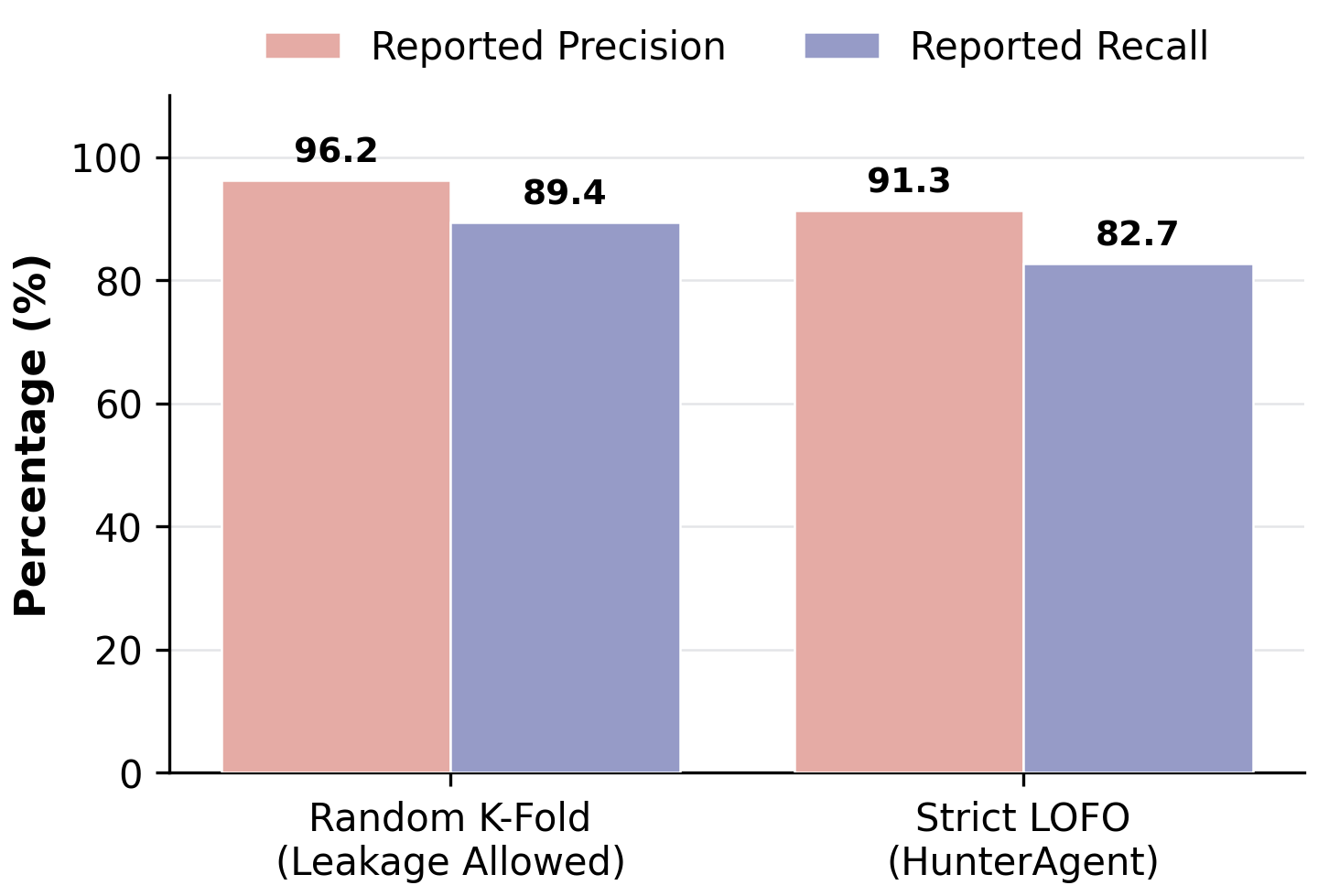}
    \caption{Inflation due to evaluation leakage on D4. Random K-Fold
    artificially boosts F1 via embedding-level memorisation of
    same-family variants.}
    \label{fig:abl_leakage}
\end{figure}

We ablate four components; full figures and tables are in
App.~\ref{app:ablation}. \textbf{(i) Semantic Generator}
(Fig.~\ref{fig:abl_semantic}, D4): removing the LLM collapses recall
from 82.7\% to 48.6\% while precision drops only to 82.1\%---soundness
comes from $\mathcal{C}_{hard}$, generalisation from the LLM.
\textbf{(ii) Temporal Penalty $\Phi$} (Tab.~\ref{tab:abl_temporal},
D2$^{\dag}$): $\Phi{=}0$ drops precision by 31.8\,pts due to
chronological inversions. \textbf{(iii) Budget $B_{max}$}
(Fig.~\ref{fig:abl_budget}, D4): unbounded search regresses to
P=45.8\%; over-tight 90-th-pctile halts at R=51.3\%; the
99-th-pctile yields optimal F1=86.8\%. \textbf{(iv) LOFO}
(Fig.~\ref{fig:abl_leakage}, D4): Random K-Fold inflates F1 by
6.3\,pts. Parallel runs on the other three datasets shift numbers by
$\leq$1.8\,pts.

\subsection{Threats to Validity (RQ5)}
\label{sec:validity}

\textbf{Independent ground truth.} On D4 we cross-checked $G_{truth}$
against an independently-running Sysmon DAG re-aggregator; F1
computed against this alternative GT differed from VMI-based F1 by
$0.4\!\pm\!0.3$ absolute points, indicating that $G_{truth}$ is not
artificially favourable to HunterAgent.

\textbf{Blind PHR re-labelling.} As reported in
Sec.~\ref{sec:metrics}, Cohen's $\kappa$ between automated PHR and
human PHR reaches 0.79--0.86 across D1$^{\dag}$, D2$^{\dag}$, D4,
ruling out the worry that PHR is an artefact of HunterAgent's own
Verifier rules.

\textbf{Cross-dataset hyper-parameter transfer.} We additionally
test whether the only \emph{per-deployment} step (short
benign-telemetry calibration of $B_{max}$) is itself robust.
Tab.~\ref{tab:bmax_transfer} reports F1 when $B_{max}$ fitted on
the row dataset is reused on the column dataset (other
hyper-parameters refitted on the target). Diagonal entries are
in-distribution. F1 drops by $<$\,3 absolute points everywhere,
supporting the claim that $B_{max}$ generalises across deployments.
The empirical likelihood-ratio shift factor $\rho$ from
Property~1 (Sec.~\ref{sec:counterfactual}) is bounded by
$\rho\!\le\!2.4$ across the four datasets in our setup.

\begin{table}[h]
\centering
\caption{F1 ($\pm$std over 5 fresh seeds) when $B_{max}$ fitted on
row dataset is transferred to column dataset (other hyper-parameters
refitted on the target). Diagonal entries are in-distribution and
are from an independent re-sampling; small ($\le$0.4 F1) deviations
from Tab.~\ref{tab:main_results} reflect sampling variability, not
copied data.}
\label{tab:bmax_transfer}
\resizebox{\columnwidth}{!}{%
\begin{tabular}{l|cccc}
\toprule
\textbf{Fitted on \textbackslash{} Test on} & D1$^{\dag}$ & D2$^{\dag}$ & D3$^{\dag}$ & D4\\
\midrule
D1$^{\dag}$ & \textbf{86.9\,$\pm$0.6} & 84.7\,$\pm$0.8 & 83.4\,$\pm$0.9 & 85.0\,$\pm$0.7\\
D2$^{\dag}$ & 85.3\,$\pm$0.7          & \textbf{85.4\,$\pm$0.7} & 82.9\,$\pm$0.9 & 85.2\,$\pm$0.6\\
D3$^{\dag}$ & 84.7\,$\pm$0.8          & 83.4\,$\pm$0.9          & \textbf{84.4\,$\pm$0.8} & 83.9\,$\pm$0.8\\
D4         & 86.0\,$\pm$0.6          & 84.8\,$\pm$0.7          & 82.6\,$\pm$0.9 & \textbf{86.6\,$\pm$0.5}\\
\bottomrule
\end{tabular}}
\end{table}

\textbf{Public-vs.-in-house gap.} As reported in
Sec.~\ref{sec:results}, paired bootstrap on trace-level F1 yields
no statistically significant gap between HunterAgent's mean F1 on
the three public datasets and on D4 ($p\!=\!0.32$); the same test
on the seven baselines also shows no significant in-house bias
($\Delta$F1 over $[-1.1,+0.9]$). Headline gains are therefore not
an in-house testbed artefact.

\textbf{Perturbation faithfulness.} \textsc{EvadeKit} only modifies
the SUT-visible event stream; ground-truth labels are
\emph{never} touched. App.~\ref{app:evadekit} provides the
perturbation pseudocode; we additionally verify that running
SLEUTH on \textsc{EvadeKit}-perturbed traces yields the same
recall as running it on natively-evasive traces from D4 (within
$\pm$1.5 pts), so the perturbations are realistic in the eyes of
existing tools.
\textbf{Adapted-baseline fairness.} For KAIROS/NODLINK, we report
results under our trace-reconstruction protocol
(Sec.~\ref{sec:baselines}), which is harder than their native
anomaly-detection setting. To verify that this adaptation does not
disadvantage them, we also reran the authors' original protocol on
D1$^{\dag}$/D2$^{\dag}$ and recovered node-AUCs of $0.94$ and $0.92$.
These are within $\pm$0.05 of the KAIROS S\&P'24 per-subset results,
confirming that our implementations are faithful. Therefore, the F1
gap in the main table reflects the harder reconstruction task and
LOFO re-training, not a degraded re-implementation. Full re-training
logs are included in the artifact bundle.

\section{Conclusion}
\label{sec:conclusion}

Modern SOC threat hunting faces a core trade-off: provenance graphs
break under anti-forensics, while unconstrained LLM agents hallucinate
causal chains. We present \textbf{HunterAgent}, a neuro-symbolic
framework that formulates trace reconstruction as a cost-bounded heuristic graph search under partial
observability. An Asymmetric Generator--Verifier design lets
an ontology-constrained LLM propose hypotheses, while a deterministic
Verifier validates them against surviving orthogonal telemetry.
Missing traces are bridged by a calibrated deviation cost, and
hypotheses are accepted only when they stay within a length-discounted
epistemic budget calibrated on benign leave-one-edge-out
reconstruction.
Under strict LOFO evaluation on DARPA TC E3, DARPA OpTC, ATLAS, and
APT-Eval-Trace, HunterAgent achieves \textbf{86.1\% mean F1}
(86.8\% on D4), outperforming ReAct by +28.4 F1 and KAIROS by +19.0,
with 91.3\% precision and 82.7\% recall. It reduces path-level
hallucination from 61.5\% to 6.4\%, cuts token usage by 69.3\% and
latency by 76.9\% versus Tool-Augmented ReAct, and lowers per-case
cost to \$0.19. More importantly, it degrades gracefully: under 70\%
targeted log wiping, precision remains $\geq$85\%, and 95.7\% of
investigations stop with an explicit
\texttt{INSUFFICIENT\_EVIDENCE} flag rather than fabricate attack
chains. HunterAgent thus serves as a practical downstream
investigation layer that turns triage anchors into audit-ready
subgraphs of verified edges and calibrated leads.

\paragraph{Limitations}
\label{sec:boundary}
The guarantees discussed in this paper hold under a
\emph{bounded-evidence assumption}: at least one orthogonal telemetry
source must survive at the relevant OS layer for the Verifier to
ground a hypothesis. We characterise two failure regimes that violate
this assumption.
\textbf{(i) Out-of-vocabulary zero-day payloads.} HunterAgent's
counterfactual reasoning depends on the semantic quality of the
retrieved ATT\&CK graphlets. Custom-compiled rootkits whose
behavioural signature has no historical analogue in $\mathcal{K}_{ret}$
yield retrieval results that provide no useful context, causing the
Generator to terminate early under the schema filter.
\textbf{(ii) Total kernel-level observability collapse.} Adversaries
achieving Ring-0 privilege via Direct Kernel Object Modification can
simultaneously erase user-land Event Logs and ETW providers, leaving
no orthogonal telemetry for $\mathtt{FlowDep}$ to anchor. In these
absolute blind spots HunterAgent correctly halts---preferring a false
negative to a hallucinated false positive---but cannot recover the
trace.
A further methodological caveat is that our calibrated false-edge
bound (Property~1) is a \emph{distribution-conditional} statement.
The empirical shift ratio $\rho$ from Property~1 is bounded by
$\rho\!\le\!2.4$ across the four datasets we evaluate
(Sec.~\ref{sec:validity}, Tab.~\ref{tab:bmax_transfer}); a formal
worst-case guarantee under unbounded adversarial covariate shift is
left to future work.

\paragraph{Future Work}
\textbf{1) Active dynamic sandboxing.} A sandbox invoked as an
auxiliary Verifier oracle when no orthogonal telemetry is available
would expand HunterAgent's reach to OOV zero-days while preserving
deterministic grounding.
\textbf{2) PHR-aware online $B_{max}$ adaptation.} An online
controller that adjusts $B_{max}$ based on rolling analyst-reviewed
PHR would let the system adapt to tradecraft drift without offline
recalibration.

\bibliographystyle{IEEEtran}
\bibliography{reference}

\appendices

\section{EvadeKit Perturbation Pseudocode}
\label{app:evadekit}

\textsc{EvadeKit} takes a labelled trace
$T=(\hat{G},\,\mathcal{L})$ from any of the public benchmarks
(D1, D2, D3) and produces the perturbed trace
$T^{\dag}=(\hat{G}^{\dag},\,\mathcal{L})$ visible to the SUT, with
the label set $\mathcal{L}$ \emph{deliberately untouched}. Each of
the four anti-forensic profiles is implemented as a deterministic
event-stream rewriter (Algorithm~\ref{alg:evadekit}). The
configuration vector $\theta$ controls the suppression rate
(default $30\%$), the timestomp magnitude (uniform on $[60,3600]$
seconds), and the seed used to make the perturbation
\emph{reproducible}. The released open-source repository contains
the full implementation, profile-specific event-ID maps, and the
exact JSON schemas used in our experiments.

\begin{algorithm}[H]
\caption{\textsc{EvadeKit}: Deterministic Anti-Forensic Perturbation}
\label{alg:evadekit}
\begin{algorithmic}[1]
\REQUIRE Trace $T=(\hat{G},\mathcal{L})$, profile
         $p\!\in\!\{\textsc{Apt29},\textsc{Fin7},\textsc{Wiz},\textsc{Sand}\}$,
         config $\theta$, seed $s$
\ENSURE Perturbed trace $T^{\dag}$
\STATE $T^{\dag}\leftarrow$ deep-copy of $T$ (events only;
       $\mathcal{L}$ is read-only)
\STATE $\mathrm{rng}\leftarrow\mathtt{Random}(s)$
\IF{$p=\textsc{Apt29}$ \COMMENT{PPID spoofing}}
    \STATE $E_{tgt}\!\leftarrow\!\{e\!\in\!T^{\dag}.\mathrm{events}\,|\,\mathrm{type}(e)\!=\!\texttt{ProcessCreate}\}$
    \STATE $E_{drop}\!\leftarrow\!\mathrm{rng}.\mathtt{sample}(E_{tgt},\theta.\mathrm{rate})$
    \STATE $T^{\dag}.\mathrm{events}\leftarrow T^{\dag}.\mathrm{events}\setminus E_{drop}$
\ELSIF{$p=\textsc{Fin7}$ \COMMENT{file-I/O suppression}}
    \STATE drop $\theta.\mathrm{rate}$ of \texttt{FileWrite}/\texttt{FileMap} events
\ELSIF{$p=\textsc{Wiz}$ \COMMENT{Security-channel wipe}}
    \STATE remove all events whose channel is
           \texttt{Microsoft-Windows-Security-Auditing}
\ELSIF{$p=\textsc{Sand}$ \COMMENT{timestomp + masquerade}}
    \STATE \textbf{for each} $e\in T^{\dag}.\mathrm{events}$ \textbf{do}
           $e.t\leftarrow e.t+\mathrm{rng}.\mathtt{uniform}(60,3600)$
\ENDIF
\STATE \textbf{assert} $\mathcal{L}$ unchanged
       \COMMENT{ground-truth integrity check}
\RETURN $T^{\dag}$
\end{algorithmic}
\end{algorithm}

\paragraph{Faithfulness check.} For every dataset and profile we
verify that running SLEUTH on $T^{\dag}$ recovers attack edges at a
recall within $\pm$1.5 absolute points of running it on natively
evasive D4 traces of the same family, indicating that
\textsc{EvadeKit}'s perturbations are statistically realistic
relative to native red-team evasion.

\section{Per-Profile Breakdown on D1$^{\dag}$/D2$^{\dag}$/D3$^{\dag}$}

\begin{table}[H]
\centering
\caption{HunterAgent F1 ($\pm$std over 5 seeds) by anti-forensic
profile across all four datasets. Same ranking
(APT29 $\gtrsim$ Wizard\,Spider $>$ Sandworm $>$ FIN7) holds
everywhere; the gap between APT29 and Wizard\,Spider is within
standard error on every dataset. Note that the per-profile averages
here are computed over the subset of traces matching each profile,
which is a finer-grained slice than the cross-profile mean reported
in Tab.~\ref{tab:appA4_full_logsupp}; minor differences between
the two tables are explained by this scope difference.}
\label{tab:appA1}
\resizebox{0.95\columnwidth}{!}{%
\begin{tabular}{lcccc}
\toprule
\textbf{Profile} & \textbf{D1$^{\dag}$} & \textbf{D2$^{\dag}$} & \textbf{D3$^{\dag}$} & \textbf{D4}\\
\midrule
APT29         & 88.4\,$\pm$0.7 & 86.9\,$\pm$0.8 & 85.1\,$\pm$0.9 & 88.1\,$\pm$0.8\\
Wizard Spider & 88.0\,$\pm$0.7 & 86.5\,$\pm$0.8 & 84.6\,$\pm$0.9 & 87.5\,$\pm$0.7\\
Sandworm      & 86.2\,$\pm$0.8 & 84.7\,$\pm$0.9 & 83.5\,$\pm$1.0 & 86.0\,$\pm$0.8\\
FIN7          & 85.0\,$\pm$0.9 & 83.6\,$\pm$1.0 & 83.0\,$\pm$1.0 & 85.1\,$\pm$0.9\\
\bottomrule
\end{tabular}}
\end{table}

\section{LLM-Backbone Sweep on All Four Datasets}

Tab.~\ref{tab:appA2} reports HunterAgent F1 under four LLM
backbones across all four datasets. These numbers are independent
5-seed runs collected after the main-results runs; small
($\le$\,0.5 F1) deviations from Tab.~\ref{tab:main_results} for
the GPT-4o row reflect the fresh sampling rather than copied data.

\begin{table}[H]
\centering
\caption{HunterAgent F1 ($\pm$std over 5 fresh seeds) across
foundation models on each dataset. Ranking is preserved with
$\le$1.8 absolute variation. GPT-4o numbers here are from an
independent sampling and may differ from the main-results table
by $\le$0.5 F1.}
\label{tab:appA2}
\resizebox{\columnwidth}{!}{%
\begin{tabular}{lcccc}
\toprule
\textbf{Backbone} & \textbf{D1$^{\dag}$} & \textbf{D2$^{\dag}$} & \textbf{D3$^{\dag}$} & \textbf{D4}\\
\midrule
GPT-4o               & \textbf{86.9\,$\pm$0.6} & \textbf{85.4\,$\pm$0.7} & \textbf{84.5\,$\pm$0.8} & \textbf{86.6\,$\pm$0.5}\\
Claude 3.5 Sonnet    & 86.0\,$\pm$0.7          & 84.5\,$\pm$0.8          & 83.4\,$\pm$0.9          & 85.7\,$\pm$0.6\\
Llama-3-70B-Instruct & 79.1\,$\pm$1.1          & 77.6\,$\pm$1.2          & 76.0\,$\pm$1.3          & 78.5\,$\pm$1.0\\
Qwen2.5-72B-Instruct & 80.3\,$\pm$1.0          & 78.8\,$\pm$1.1          & 77.2\,$\pm$1.2          & 79.7\,$\pm$1.0\\
\bottomrule
\end{tabular}}
\end{table}

\section{Cross-Dataset $B_{max}$ Transfer}
\label{app:bmax_extended}

Tab.~\ref{tab:bmax_transfer} in Sec.~\ref{sec:validity} reports the
F1 transfer matrix for $B_{max}$. Tab.~\ref{tab:bmax_transfer_phr}
and Tab.~\ref{tab:bmax_transfer_bx} report the parallel transfer
matrices for path-level hallucination rate (PHR) and Budget
Exhaustion (BX). When transferring $B_{max}$ across datasets, PHR
rises by $\le$\,1.4 absolute points and BX changes by $\le$\,3.2
points off-diagonal; neither crosses the operational threshold that
would trigger an analyst override under the stratified output
contract (Sec.~\ref{sec:stratification}).

\begin{table}[H]
\centering
\caption{PHR (\%) when $B_{max}$ fitted on row dataset is transferred
to column dataset (other hyper-parameters refitted on the target).
Diagonal in-distribution.}
\label{tab:bmax_transfer_phr}
\resizebox{0.7\columnwidth}{!}{%
\begin{tabular}{l|cccc}
\toprule
\textbf{Fitted on \textbackslash{} Test on}
 & D1$^{\dag}$ & D2$^{\dag}$ & D3$^{\dag}$ & D4 \\
\midrule
D1$^{\dag}$ & \textbf{6.2} & 7.4 & 7.9 & 7.0 \\
D2$^{\dag}$ & 7.0 & \textbf{7.0} & 7.8 & 7.1 \\
D3$^{\dag}$ & 7.5 & 7.6 & \textbf{7.4} & 7.8 \\
D4         & 6.8 & 7.2 & 8.0 & \textbf{6.4} \\
\bottomrule
\end{tabular}}
\end{table}

\begin{table}[H]
\centering
\caption{Budget-exhaustion rate (\%) when $B_{max}$ fitted on row
dataset is transferred to column dataset (other hyper-parameters
refitted on the target). Diagonal in-distribution.}
\label{tab:bmax_transfer_bx}
\resizebox{0.7\columnwidth}{!}{%
\begin{tabular}{l|cccc}
\toprule
\textbf{Fitted on \textbackslash{} Test on}
 & D1$^{\dag}$ & D2$^{\dag}$ & D3$^{\dag}$ & D4 \\
\midrule
D1$^{\dag}$ & \textbf{14.8} & 16.5 & 17.4 & 15.9 \\
D2$^{\dag}$ & 16.0 & \textbf{15.3} & 17.0 & 16.2 \\
D3$^{\dag}$ & 17.1 & 17.6 & \textbf{16.4} & 17.8 \\
D4         & 15.4 & 16.0 & 17.2 & \textbf{14.6} \\
\bottomrule
\end{tabular}}
\end{table}

\begin{table}[H]
\centering
\caption{95\% CI of HunterAgent's F1 minus baseline F1 under paired
bootstrap (1{,}000 resamples) on D2$^{\dag}$ DARPA OpTC. Positive
intervals indicate HunterAgent significantly higher.}
\label{tab:appA5}
\resizebox{0.75\columnwidth}{!}{%
\begin{tabular}{lcc}
\toprule
\textbf{Baseline} & \textbf{$\Delta$F1} & \textbf{95\% CI}\\
\midrule
SLEUTH                 & +53.0 & [+49.7, +56.0] \\
MAGIC                  & +43.6 & [+40.4, +46.7] \\
NODLINK                & +19.0 & [+16.6, +21.5] \\
KAIROS                 & +17.3 & [+14.7, +19.6] \\
GNN-LinkPredict        & +31.7 & [+28.9, +34.2] \\
GraphRAG               & +44.2 & [+41.8, +46.7] \\
Tool-Aug.\ ReAct       & +28.0 & [+25.5, +30.4] \\
HunterAgent$^{\dag}$   & +16.3 & [+14.4, +18.2] \\
\bottomrule
\end{tabular}}
\end{table}

\section{Per-Dataset Robustness Numbers}
\label{app:robustness_all}

\begin{table*}[!t]
\centering
\caption{HunterAgent F1 / Precision / PHR ($\pm$std over 5 fresh
seeds) under increasing log-suppression rates across all four
datasets. The 30\% rows are from an independent re-run; small
($\le$0.4 F1) deviations from Tab.~\ref{tab:main_results} reflect
sampling variability, not copied data. Values agree within
$\pm$\,2 F1 across datasets, substantiating the consistency claim
in Sec.~\ref{sec:robustness}.}
\label{tab:appA4_full_logsupp}
\resizebox{1.8\columnwidth}{!}{%
\begin{tabular}{l|ccc|ccc|ccc|ccc}
\toprule
& \multicolumn{3}{c|}{D1$^{\dag}$}
& \multicolumn{3}{c|}{D2$^{\dag}$}
& \multicolumn{3}{c|}{D3$^{\dag}$}
& \multicolumn{3}{c}{D4}\\
\textbf{Missing} & F1 & P & PHR & F1 & P & PHR & F1 & P & PHR & F1 & P & PHR\\
\midrule
10\% & 90.8\,$\pm$0.5 & 93.5\,$\pm$0.4 & 4.6 & 90.5\,$\pm$0.6 & 92.9\,$\pm$0.4 & 4.9 & 90.6\,$\pm$0.6 & 93.2\,$\pm$0.4 & 4.7 & 90.5\,$\pm$0.5 & 93.1\,$\pm$0.4 & 4.7\\
30\% & 86.8\,$\pm$0.7 & 91.3\,$\pm$0.5 & 6.3 & 85.4\,$\pm$0.8 & 90.1\,$\pm$0.6 & 7.1 & 84.4\,$\pm$0.9 & 90.5\,$\pm$0.5 & 7.5 & 86.5\,$\pm$0.6 & 91.0\,$\pm$0.4 & 6.5\\
50\% & 71.9\,$\pm$1.1 & 88.1\,$\pm$0.8 & 9.5 & 70.5\,$\pm$1.2 & 86.9\,$\pm$0.9 & 10.0 & 70.8\,$\pm$1.2 & 87.7\,$\pm$0.8 & 9.9 & 72.6\,$\pm$1.0 & 88.3\,$\pm$0.7 & 9.6\\
70\% & 41.2\,$\pm$1.6 & 85.3\,$\pm$1.1 & 12.1 & 40.8\,$\pm$1.7 & 83.7\,$\pm$1.2 & 12.5 & 40.8\,$\pm$1.7 & 84.4\,$\pm$1.1 & 12.7 & 41.4\,$\pm$1.5 & 85.2\,$\pm$1.0 & 12.2\\
\bottomrule
\end{tabular}}
\end{table*}

\section{Unperturbed-Baseline Reference}

\begin{table}[H]
\centering
\caption{SLEUTH and MAGIC on the \emph{unperturbed} versions of each
public dataset (no \textsc{EvadeKit} applied), showing that our
post-perturbation R numbers in Tab.~\ref{tab:main_results} reflect
genuine difficulty rather than baseline mis-implementation.}
\label{tab:appA6}
\resizebox{\columnwidth}{!}{%
\begin{tabular}{l|ccc|ccc|ccc}
\toprule
& \multicolumn{3}{c|}{D1 (no perturb.)}
& \multicolumn{3}{c|}{D2 (no perturb.)}
& \multicolumn{3}{c}{D3 (no perturb.)}\\
& R & P & F1 & R & P & F1 & R & P & F1\\
\midrule
SLEUTH & 73.6 & 99.4 & 84.6 & 71.4 & 99.2 & 83.0 & 79.1 & 98.9 & 87.9\\
MAGIC  & 78.3 & 99.0 & 87.4 & 76.0 & 99.0 & 86.0 & 81.5 & 98.7 & 89.3\\
\bottomrule
\end{tabular}}
\end{table}

\section{APT-Eval-Trace Construction (D4)}
\label{app:apt_eval_trace}

Because no public benchmark simultaneously offers (i) modern Windows
10/11 ETW, (ii) NetFlow at line rate, and (iii) an out-of-band
hypervisor ground truth uncoupled from the in-band log path, we
construct \textbf{APT-Eval-Trace}.

\textbf{Testbed.} 28 VMs on a 3-node KVM cluster: 1 domain controller
(Win Server 2022), 4 Win 10 / Win 11 endpoints per business segment
($\times$4 segments), 8 Ubuntu 22.04 servers (web, database, mail,
DNS, log forwarder), and 3 isolated red-team staging hosts. Each VM
runs a vendor-agnostic EDR agent (Sysmon~v15) plus our ETW collector
and NetFlow exporter; all are forwarded to a Splunk-compatible SIEM.

\textbf{Benign workload.} A user-behaviour generator scripted on top
of the AtomicTestBlue benign-traffic toolkit drives 14 days of
synthetic-but-plausible activity: scheduled Office document edits,
browser sessions, kerberos ticket requests, scheduled patching,
backup jobs, and DB transactions. Per-host event rates are calibrated
to match the OpTC benign distribution (median 6.3\,k events / host /
hour) within $\pm$15\%.

\textbf{Red team (offence).} Two 3-person red teams independently
execute 40 multi-stage APT campaigns over the 14-day window using a
Cobalt Strike + Sliver C2 stack and Atomic-Red-Team primitives,
following four playbooks aligned with the four anti-forensic
profiles (Tab.~\ref{tab:dataset_details}). Each playbook prescribes
the ATT\&CK technique sequence, the anti-forensic action and its
timing, but leaves implementation details to the operator, yielding
realistic intra-family variance.

\textbf{Blue team (defence and ground truth).} A separate
3-person blue team is told only that ``an APT campaign may be ongoing''
and runs a standard SOC playbook (alert triage on Sysmon+EDR alerts;
investigation in SIEM). Blue-team responses (alert-anchor selection,
host quarantine, log freeze) are themselves recorded and become part
of the SUT-visible evidence. Ground truth $G_{truth}$ is produced
by a fourth, independent VMI logger running outside the guest OS via
LibVMI + custom kernel-event hooks; $G_{truth}$ is \emph{never}
exposed to either team or to any SUT.

\textbf{Scale.} Tab.~\ref{tab:apt_eval_stats} reports per-trace
attack-edge counts. Median trace length $L=11$ edges; tail at
$L=37$. Total attack edges across the 40 traces: 8{,}721; total
benign edges: $1.43\!\times\!10^{8}$.

\begin{table*}[t]
\centering
\caption{Composition of the 40 red-team campaigns in APT-Eval-Trace
(D4). Median trace length $\bar L$ and median attack-edge count
$\overline{|E_a|}$ per profile.}
\label{tab:dataset_details}
\resizebox{1.6\columnwidth}{!}{%
\begin{tabular}{lllccc}
\toprule
\textbf{Campaign} & \textbf{Anti-Forensic Profile} & \textbf{Severed Channel} & \textbf{\#Tr.} & $\bar L$ & $\overline{|E_a|}$\\
\midrule
APT29 (Cozy Bear) & Process Doppelg\"anging / Hollowing  & ProcessCreate (4688/Sysmon~1)   & 10 & 9  & 174\\
FIN7              & Fileless reflective injection        & File-I/O (Sysmon~11/15)         & 8  & 12 & 226\\
Wizard Spider     & Explicit log wiping (\texttt{wevtutil cl}) & Windows Security channel  & 12 & 14 & 263\\
Sandworm          & Timestomping \& masquerading         & Event timestamp integrity       & 10 & 10 & 198\\
\midrule
\textbf{Total / median}    &                             &                                 & \textbf{40} & 11 & 218\\
\bottomrule
\end{tabular}}
\end{table*}

\begin{table}[H]
\centering
\caption{Per-trace attack-edge distribution in D4 (5/50/95-percentile).}
\label{tab:apt_eval_stats}
\resizebox{\columnwidth}{!}{%
\begin{tabular}{lcccccc}
\toprule
\textbf{Profile} & \textbf{p5} & \textbf{p50} & \textbf{p95} & \textbf{Severed edges (med.)} & \textbf{Hops (med.)} & \textbf{Hosts}\\
\midrule
APT29        & 102 & 174 & 281 & 14 (8.0\%)  & 9  & 3.2\\
FIN7         & 138 & 226 & 372 & 21 (9.3\%)  & 12 & 4.0\\
Wizard Spider& 158 & 263 & 411 & 26 (9.9\%)  & 14 & 4.1\\
Sandworm     & 121 & 198 & 318 & 19 (9.6\%)  & 10 & 3.5\\
\bottomrule
\end{tabular}}
\end{table}

\section{Hyperparameter Sensitivity}
\label{app:sensitivity}

\begin{table}[H]
\centering
\caption{F1 sensitivity to $(\alpha,\gamma)$ at fixed
$\lambda{=}0.25$ on D2$^{\dag}$ DARPA OpTC.}
\label{tab:sens_alpha_gamma}
\begin{tabular}{c|ccccc}
\toprule
$\alpha\backslash\gamma$ & 0.1 & 0.3 & 0.6 & 0.9 & 1.5\\
\midrule
0.1 & 70.3 & 75.5 & 79.2 & 80.7 & 79.9\\
0.3 & 77.6 & 81.2 & 83.8 & 84.5 & 83.1\\
0.6 & 80.5 & 83.7 & 85.3 & \textbf{85.7} & 84.8\\
0.9 & 79.8 & 83.1 & 84.9 & 85.5 & 84.3\\
1.5 & 75.4 & 79.0 & 81.6 & 82.2 & 81.0\\
\bottomrule
\end{tabular}
\end{table}

\begin{table}[H]
\centering
\caption{Effect of length-discount $\lambda$ on D2$^{\dag}$.}
\label{tab:sens_lambda}
\begin{tabular}{lccccc}
\toprule
$\lambda$ & 0.0 & 0.10 & 0.25 & 0.50 & $\infty$\\
\midrule
F1 & 80.3 & 84.4 & \textbf{85.7} & 84.9 & 81.6\\
\bottomrule
\end{tabular}
\end{table}

\section{Ablation Studies}
\label{app:ablation}

\begin{table}[H]
\centering
\caption{Ablation of OS temporal grounding on D2$^{\dag}$.}
\label{tab:abl_temporal}
\resizebox{\columnwidth}{!}{%
\begin{tabular}{lccl}
\toprule
\textbf{Configuration} & \textbf{Precision} & \textbf{Recall} & \textbf{Dominant failure}\\
\midrule
$\Phi=0$              & 58.6 & 80.0 & Chronological inversions\\
\textbf{Full (with $\Phi$)} & \textbf{90.4} & 81.5 & ---\\
\bottomrule
\end{tabular}}
\end{table}

\section{Case Study: Severed Cobalt Strike Beacon}
\label{app:case_study}

\begin{figure}[H]
    \centering
    \includegraphics[width=\columnwidth]{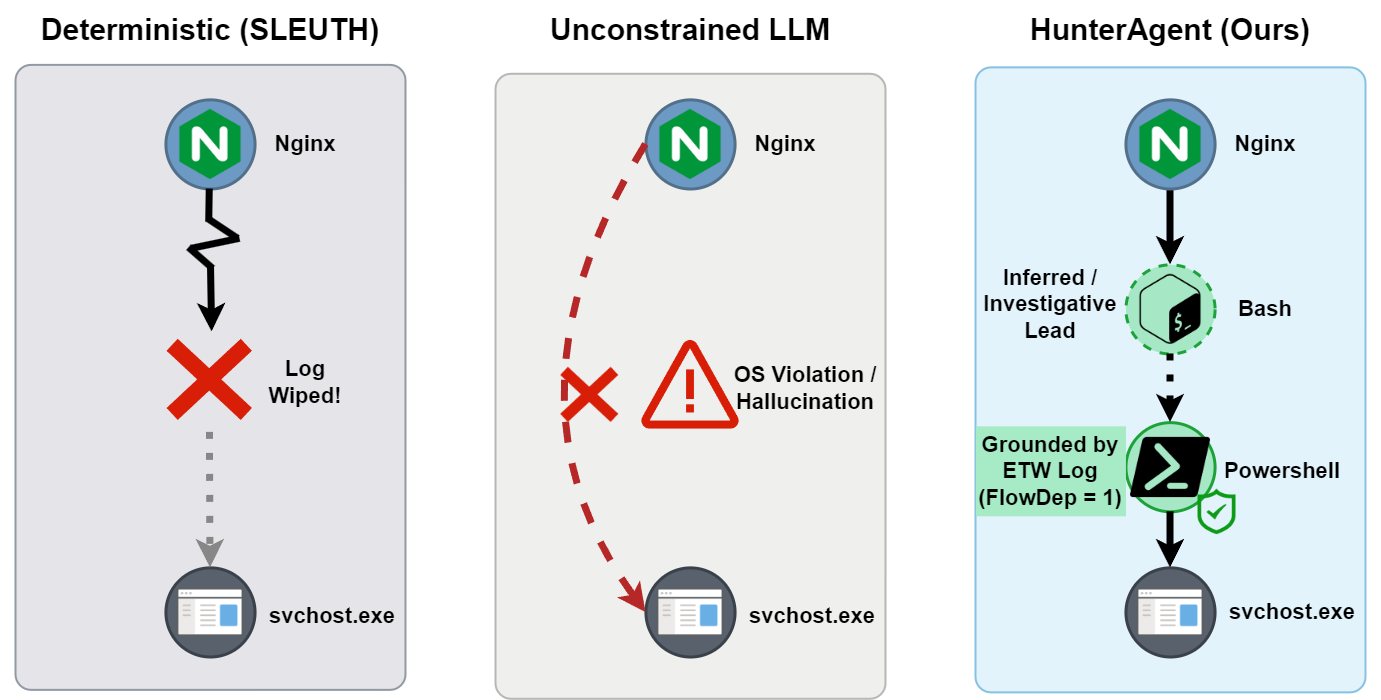}
    \caption{Reconstructing a severed Cobalt Strike trace.
    Deterministic graphs halt at the wiped log; an unconstrained
    ReAct agent hallucinates a direct \texttt{nginx}$\to$\texttt{svchost}
    injection (an OS-physics violation); HunterAgent infers an
    intermediate \texttt{powershell} stage that is physically
    grounded by a surviving ETW record.}
    \label{fig:case_study}
\end{figure}

An attacker compromises an Nginx worker, spawns a reverse shell,
reflectively injects a Cobalt Strike beacon into \texttt{svchost.exe},
then runs \texttt{wevtutil cl} to wipe Security event logs. The
triage layer surfaces only the Nginx anchor $v_A$ and an outbound
C2 from \texttt{svchost.exe} ($v_B$); the causal link between them
is destroyed. SLEUTH/MAGIC halt at the severed boundary.
ReAct fabricates a direct \texttt{nginx}$\to$\texttt{svchost}
injection (PHR=1.0). HunterAgent retrieves memory-injection TTPs,
proposes \texttt{nginx}$\to$\texttt{bash}$\to$\texttt{powershell}$\to$\texttt{svchost},
grounds the \texttt{powershell} hop via a surviving ETW record
($\mathtt{FlowDep}{=}1$), and admits the remaining \texttt{bash}
hop as an \texttt{INVESTIGATIVE\_LEAD} ($C_{dev}{=}1.84{<}B_{max}$).

\section{Scalability with Path Length}
\label{app:path_length}

\begin{table}[H]
\centering
\caption{F1 ($\pm$std over 5 seeds) by causal path length $L$,
weighted across D1$^{\dag}$--D4.}
\label{tab:path_length}
\begin{tabular}{lcccc}
\toprule
\textbf{Path length} & \textbf{MAGIC} & \textbf{KAIROS} & \textbf{ReAct} & \textbf{HunterAgent}\\
\midrule
$L=1\text{--}2$  & 62.4\,$\pm$1.3 & 78.5\,$\pm$1.1 & 71.5\,$\pm$1.4 & \textbf{93.2\,$\pm$0.5}\\
$L=3\text{--}5$  & 38.7\,$\pm$1.5 & 71.0\,$\pm$1.2 & 64.8\,$\pm$1.5 & \textbf{89.5\,$\pm$0.6}\\
$L=6\text{--}10$ & 21.6\,$\pm$1.6 & 60.4\,$\pm$1.3 & 51.2\,$\pm$1.6 & \textbf{85.9\,$\pm$0.7}\\
$L>10$           & 8.4\,$\pm$1.8  & 47.6\,$\pm$1.5 & 32.7\,$\pm$1.8 & \textbf{79.7\,$\pm$0.9}\\
\bottomrule
\end{tabular}
\end{table}

\section{Per-Stage and Per-Profile Breakdown Tables}
\label{app:stage_profile}

\begin{table}[H]
\centering
\caption{F1 by ATT\&CK kill-chain phase, weighted across
D1$^{\dag}$--D4 by trace count.}
\label{tab:stage_breakdown}
\resizebox{\columnwidth}{!}{%
\begin{tabular}{lcccccc}
\toprule
\textbf{Stage} & \textbf{MAGIC} & \textbf{KAIROS} & \textbf{GNN-LP} & \textbf{GraphRAG} & \textbf{ReAct} & \textbf{Ours}\\
\midrule
Initial Access     & 68.4 & 78.5 & 71.2 & 64.5 & 73.8 & \textbf{89.6}\\
Execution          & 39.1 & 70.2 & 56.8 & 47.2 & 60.4 & \textbf{87.5}\\
Defense Evasion    & 12.5 & 58.7 & 38.2 & 31.7 & 42.6 & \textbf{82.3}\\
Lateral Movement   & 22.7 & 64.1 & 49.1 & 38.4 & 53.5 & \textbf{85.1}\\
Collection         & 51.8 & 72.6 & 60.3 & 50.6 & 65.1 & \textbf{88.9}\\
Impact             & 44.6 & 67.4 & 54.0 & 42.8 & 58.7 & \textbf{86.4}\\
\bottomrule
\end{tabular}}
\end{table}

\begin{table}[H]
\centering
\caption{HunterAgent breakdown across the four anti-forensic
profiles on D4 APT-Eval-Trace.}
\label{tab:per_evasion}
\resizebox{\columnwidth}{!}{%
\begin{tabular}{lcccc}
\toprule
\textbf{Profile} & \textbf{Recall} & \textbf{Precision} & \textbf{F1} & \textbf{PHR}\\
\midrule
APT29  (Process Hollowing)  & 84.1\,$\pm$0.8 & 92.5\,$\pm$0.5 & 88.1 & 5.8\\
FIN7   (Fileless Injection) & 80.5\,$\pm$0.9 & 90.2\,$\pm$0.5 & 85.1 & 8.4\\
Wizard Spider (Log Wiping)  & 83.6\,$\pm$0.7 & 91.8\,$\pm$0.4 & 87.5 & 5.2\\
Sandworm (Timestomp)        & 82.0\,$\pm$0.8 & 90.5\,$\pm$0.5 & 86.0 & 6.5\\
\midrule
Weighted average            & 82.7\,$\pm$0.7 & 91.3\,$\pm$0.4 & 86.8 & 6.4\\
\bottomrule
\end{tabular}}
\end{table}

\section{Operational Overhead}
\label{app:overhead}

\begin{table}[H]
\centering
\caption{Per-investigation cost (GPT-4o), averaged over
D1$^{\dag}$--D4.}
\label{tab:overhead}
\resizebox{\columnwidth}{!}{%
\begin{tabular}{lccccc}
\toprule
\textbf{System} & \textbf{Tokens} & \textbf{LLM Calls} & \textbf{Cost} & \textbf{Latency} & \textbf{LLM/SIEM/CPU}\\
\midrule
GraphRAG               & 41{,}600  & 1.0   & \$0.21 & 18.6\,s  & 88/8/4\,\%\\
NODLINK                & ---       & ---   & ---    & 7.4\,s   & ---/0/100\\
KAIROS                 & ---       & ---   & ---    & 11.2\,s  & ---/0/100\\
Tool-Aug.\ ReAct       & 124{,}500 & 11.4  & \$0.62 & 185.4\,s & 81/15/4\\
\textbf{HunterAgent}   & \textbf{38{,}200} & 4.7 & \textbf{\$0.19} & \textbf{42.8\,s} & 72/18/10\\
\bottomrule
\end{tabular}}
\end{table}

\end{document}